\documentclass{CUP-JNL-DTM}%

\usepackage{graphicx}
\usepackage{multicol,multirow}
\usepackage{amsmath,amssymb,amsfonts}
\usepackage{mathrsfs}
\usepackage{amsthm}
\usepackage{rotating}
\usepackage{appendix}
\usepackage{float}  

\usepackage{tikz}
\usepackage{subcaption}  
\usepackage{enumitem} 

\usepackage[natbib,style=apa]{biblatex}
\usepackage{ifpdf}
\usepackage[T1]{fontenc}
\usepackage{newtxtext}
\usepackage{newtxmath}
\usepackage{textcomp}
\usepackage{xcolor}
\usepackage{lipsum}
\usepackage[colorlinks,allcolors=blue]{hyperref}

\theoremstyle{definition}

\numberwithin{equation}{section}

\jissue{1}
\begin{document}

\setlength{\abovedisplayskip}{0pt}  
\setlength{\belowdisplayskip}{4pt}
\setlength{\abovedisplayshortskip}{0pt}
\setlength{\belowdisplayshortskip}{4pt}

\begin{Frontmatter}

\title[Article Title]{Crafting desirable climate trajectories with RL explored socio-environmental simulations}

%

\author[1]{James Rudd-Jones}
\author[1]{Fiona Thendean}
\author[1]{María Pérez-Ortiz}

\authormark{James Rudd-Jones \textit{et al}.}

\address[1]{\orgdiv{UCL Centre for Artificial Intelligence, Department of Computer Science}, \orgname{University College London}, \orgaddress{\city{London}, \country{United Kingdom}}}


\authormark{James Rudd-Jones et al.}



\abstract{

Climate change poses an existential threat, necessitating effective climate policies to enact impactful change. Decisions in this domain are incredibly complex, involving conflicting entities and evidence. In the last decades, policymakers increasingly use simulations and computational methods to guide some of their decisions. Integrated Assessment Models (IAMs) are one of such methods, which combine social, economic, and environmental simulations to forecast potential policy effects. For example, the UN uses outputs of IAMs for their recent Intergovernmental Panel on Climate Change (IPCC) reports. Traditionally these have been solved using recursive equation solvers, but have several shortcomings, e.g. struggling at decision making under uncertainty. Recent preliminary work using Reinforcement Learning (RL) to replace the traditional solvers shows promising results in decision making in uncertain and noisy scenarios. We extend on this work by introducing multiple interacting RL agents as a preliminary analysis on modelling the complex interplay of socio-interactions between various stakeholders or nations that drives much of the current climate crisis. Our findings show that cooperative agents in this framework can consistently chart pathways towards more desirable futures in terms of reduced carbon emissions and improved economy. However, upon introducing competition between agents, for instance by using opposing reward functions, desirable climate futures are rarely reached. Modelling competition is key to increased realism in these simulations, as such we employ policy interpretation by visualising what states lead to more uncertain behaviour, to understand algorithm failure. Finally, we highlight the current limitations and avenues for further work to ensure future technology uptake for policy derivation.

}
\end{Frontmatter}

\section*{Impact Statement}


Deriving climate policy is a challenging problem, with an expansive solution space.
Policymakers have turned to simulation based approaches in order to aid their decisions, however these traditionally have various limitations. 
Our work is a preliminary study on improving aspects of these simulation based approaches with multi-entity agent interactions.
This allows for improved modelling of stakeholder/nation competition, cooperation, and communication that is the key driver for much of anthropogenic climate change.

\section{Introduction}

According to the 2022 Intergovernmental Panel on Climate Change (IPCC) report - ``\emph{Having the right policies, infrastructure and technology in place to enable changes to our lifestyles and behaviour can result in a 40-70\% reduction in greenhouse gas emissions by 2050}" \citep{Theevide7:online}. 
The overall findings show that within all sectors technology exists that will enable a habitable future, but their adoption may require capital intensive investments, and societal changes. 
Ambitious policies can have some effect on incentivising funding towards research or implementation of such technologies, and enforcing certain behavioural restrictions, but are not the exclusive driver to change lifestyle and behaviour.
These major policy change adjustments which are needed to combat climate change, can therefore be met with strong opposition that prevents uptake \citep{patterson2023backlash}, as entrenched societal structures, cultural norms, and vested interests often resist shifts that challenge the status quo.
Evidence based policy is key here as it not only improves the derived policy but states quantifiable results that can reassure critics \citep{cairney2016politics}. However, this can be challenging within the climate domain as we are experiencing novel events that have never been tackled. Climate modelling through simulations greatly helps as it provides evidence of future trajectories and attributes metrics to how future actions can have an impact. With human behaviour so inextricably linked to our changing climate it is key that these simulation models incorporate human factors to not exclude anthropogenic effects. 
Models of this type are known as Integrated Assessment Models (IAMs), that join traditional climate simulations with socio-economic dynamic models \citep{dowlatabadi1995integrated}. The UN extensively uses outputs of IAMs for the backbone of their IPCC reports, submitted by researchers across the world, providing quantitative insights into the trade-offs and synergies between different policy options and their consequences on socio-economic and/or environmental factors \citep{van2020anticipating}. On the UN website they publicly list twenty-nine IAMs used for their decision making \citep{Integrat69:online}, such as the GEMINI-E3 model that specifically assesses how world climate change policies affect countries both at the micro and macro economic levels \citep{bernard2008gemini}. 
As an example \cite{van2023multimodel} use multiple IAMs to analyse how the national policies and pledges made at the latest COP26 Glasgow conference will affect future $CO_2$ emission trajectories, one of which being GEMINI-E3.

IAMs are the current most used model framework for the socio-environmental domain, traditionally paired with an optimal control problem (for example Model Predictive Control \citep{garcia1989MPC}), to predict future trajectories towards a desired outcome \citep{kellett2019feedback}.
However they are not free from their own shortcomings. 
Some key negatives are their poor representation of behavioural and economic systems as well as a lack of modelling decision-making under uncertainty \citep{farmer2015third, zhang2022ai}, for further details refer to the review of \cite{gambhir2019review}. Both can be improved using Agent-Based Model (ABM) approaches \citep{gambhir2019review}. ABMs are a common within domains such as financial modelling \citep{axtell2022agent} or transport modelling \citep{wise2017transportation} as they allow agent heterogeneity, agent cooperation/competition/communication, closer representative entity dynamics to reality, and more \citep{axtell2022agent}. 
These features improve decision-making over the traditional control problem, but require agent behavioural policies to be defined (rather than learnt) outside of the simulation, which can still struggle under uncertainty \citep{kelly1999integrated, van2019improved}. Further improvements on ABMs incorporate trained algorithms to infer the best actions and search the solution space instead of heuristic behavioural policies. This deeper exploration increases an agent's robustness to simulation uncertainty, which is paramount with the highly changeable simulation dynamics caused by the current climate. In this case ABMs must be reformulated so that agents receive a signal (e.g. a reward) from the environment after each action taken, that is used to update their behavioural policy.




Reinforcement Learning (RL) and especially Multi-Agent Reinforcement Learning (MARL) algorithms are widely used within ABM literature to improve agent behaviour policies \citep{sert2020segregation, liang2020agent}. We carry this RL theme over, replacing the control problem on top of the IAM environment simulation to increase exploration in this space. Temporally updating agents account for the changeability in the climate simulations caused by their own and other agent's actions, creating feedback loops that enable reactive behaviour to further climate or other agent changes. 
Another benefit of this MARL approach is that it is simulation agnostic, extended developments in the field can be applied to any form of multiple agent simulation be it IAMs, ABMs, etc, although would require further training.


The application of RL and MARL to IAMs is a novel topic with only a handful of previous works. For a single agent scenario, the work of \cite{strnad2019deep} and our previous work in \cite{wolf2022climate} applied an RL agent into an IAM, that once trained was able to generate policy guidance pathways towards a defined ``economic and environmental positive future" within the models framework. They focused on adapting agent initial states and reward functions to understand the impact these had on the exploration of the IAM, as well as test the agents under the injection of noise in the environment. This has guided our experiments to ensure a wide range of initialisations to understand the exploration of agents. Both \cite{strnad2019deep} and our previous work in \cite{wolf2022climate} use a singular agent, hence assuming a ''unified" earth, in which there is a collectively shared goal. In this work we aim to move one step further and model inter-world interactions, that are the driver for much of anthropogenic climate change and must be understood for many policy decisions \citep{stone2008global}. Towards this aim we adapt the IAM accordingly, based on ABM extensions of IAMs \citep{nordhaus2015climate, giarola2022muse, zhang2022ai}, in order to implement a multi-agent IAM with MARL.

The only work that has used MARL within the climate policy domain in the literature is that of \cite{zhang2022ai}, which created the RICE-N model used for the AI for Global Climate Cooperation Challenge\footnote{AI For Global Climate Cooperation competition - \protect\url{https://www.ai4climatecoop.org}}. 
Itself an extension of the Regional Integrated model of Climate and the Economy (RICE) model developed in \cite{nordhaus2010economic} that models twelve global regions.
\cite{zhang2022ai} invited various domain experts to create and edit interaction and negotiation protocols to achieve the best Pareto Frontier of the socio-economic system variables in the environment.
The RICE-N model combines a climate-economic IAM with trade and negotiation dynamics enabling high levels of interaction between countries/regions (a.k.a agents). Agents can adjust their savings rates, climate mitigation rates, as well as trade and negotiate with each other at each time step, leading to a large range of potential interactions between each other and the environment  \citep{zhang2022ai}. Their findings show the potential of MARL based applications to IAMs with a large call to action for further research on the topic. RICE-N is an extensive environment that we aim to use for future work, however we prioritise increased intepretability of the trained agent and as such focus on the multi-agent extension of the more simplistic environment as used in \cite{strnad2019deep} and \cite{wolf2022climate}. This simplified environment enables a visual understanding and easier interpretation of the trained agent's interactions, which are key to analyse the use of MARL within IAMs. 

RL algorithms however, lack inherent explainability, raising concerns about their trustworthiness for informing real-world policy decisions. Using explainability methods, we can reinforce human confidence by providing insights into how decisions were made and visibility to vulnerabilities \citep{Adadi2018, Lipton2018, Glanois2021}. The explainability methods explored in this work specifically target explaining model policy through a quantification technique, determining the states at which taking a certain action is crucial, critical in applications related to informing climate change policy.

In summary, we attempt to model whether agents prioritising economic or environmental gain can affect climate policy derivation. As well as simulate, within this framework, whether ``climate positive" futures are possible when agents conflict in their prioritisations.
We have extended previous literature's single agent IAM to a multi-agent scenario in order to incorporate inter-nation behaviour.
Utilising this technology, policies can be derived and enacted in reality, depending on the validity of our underlying IAM.
For a single agent setting, one can fully implement the projected policies as they can have full agency over the singular agent in reality.
However, moving to multiple agents if we want to follow a similar optimisation approach it assumes we can have control over all agents in reality. 
A heavy assumption in practice.
Instead in this paper we focus on the setting of having control over one or a subset of the agents, but still model all agents learning collectively. 
This necessitates the need for decentralised training decentralised execution (DTDE) algorithms.
We have arbitrarily assumed the learning algorithm and parameters behind each stylised agent, which will directly affect the outcome trajectories.
Aiming to highlight the challenges with employing certain existing algorithms.
However in future work, the other agents in the simulation (that we may not have agency over in reality) could be trained using imitation learning \citep{hussein2017imitation} on historical data to represent in-silico versions of real world entities.
MARL can then be used to train an agent to act as a best response to these imitation pre-trained agents within a multiple agent IAM, providing us with a range of possible future trajectories. Again dependent on the validity not only of the IAM, but also the agent representations of real world entities.
As with any forecasting tool, long range trajectories lead to large accumulations of error. As an alternative the algorithm can be further trained as more data about other agents is received.
Finally an inherent challenge with algorithm derived policy is being able to interpret the underlying solution, especially in edge cases or failure scenarios in which there may not be much prior experience. We have implemented initial interpretability techniques to increase trust in the system for down stream applications.

Our results show that multiple agents that work towards the same goal cooperatively are able to achieve the IAMs ``economic and environmental positive future" success state consistently over 90\% of test episodes. Increasing competition between agents reduces this success significantly, which is one of this work's main conclusions, and is a major avenue for future work, as in reality competition or mixed motivations are rife. This work places as an early discovery into the field positioning future research required to achieve adoption of the technology. The code to run our experiments will be publicly available online upon acceptance of the manuscript.



\section{Materials and methods}

In this section, we introduce the core themes required for our contribution: the IAM environment, the MARL algorithm and requirements for its application, and the interpretability framework we have used in order to improve insight.

\subsection{The IAM Environment}
The AYS environment, created by \cite{kittel2021lakes}, is a low complexity IAM, made up of a social, economic, and environmental variable. These three variables each relate to an ordinary differential equation (ODE) defining the system:
\begin{align}
    \frac{dA}{dt} &= E - \frac{A}{\tau_A} \\
    \frac{dY}{dt} &= \beta Y - \theta A Y \\
    \frac{dS}{dt} &= R - \frac{S}{\tau_S}
\end{align}

where $A$ is the excess atmospheric carbon ($GtC$), $Y$ the economic output ($\text{\$yr}^{-1}$), and $S$ the renewable knowledge stock ($GJ$). Each variable is inextricably linked with each other, creating a dynamic cycle. In words:

\begin{itemize}
    \item $A$ is proportional to emissions produced from the use of fossil fuels, minus a natural carbon decay out of the atmosphere.
    \item $Y$ naturally grows by 3\% each time period however, is reduced by a economic climate damage function where increasing $A$ increases the reduction in $Y$.
    \item $S$ is proportional to the amount of renewable energy produced, however, has a natural knowledge decay rate over time.
\end{itemize}

The following equations are required for deeper analysis of the AYS ODEs, with further numerical parameters listed in Appendix \ref{appendix:further_ays_details}.

\begin{align}
    &\text{Emissions} \quad \quad \quad \quad \quad \quad \quad \quad \quad \; \; E = \frac{\Gamma U}{\phi} \label{eqn:emissions} \\
    &\text{Fossil Fuel Energy Share} \quad \quad \quad \; \Gamma = \frac{1}{1 + (\frac{S}{\sigma}) ^ \rho} \\
    &\text{Energy Demand} \quad \quad \quad \quad \quad \quad \quad U = \frac{Y}{\epsilon} \\
    &\text{Renewable Energy Produced} \quad R = (1 - \Gamma) U 
\end{align}


Whilst A and Y are easily quantifiable with real life implications, S is harder to define. Generally social factors require greater levels of detail than economic or environmental attributes. For instance, in \cite{zhang2022ai} they incorporate many layers of complex socio-economic equations in order to have a functioning model with quantifiable social impact. 
In the AYS model this is simplified down to a single equation enabling a much reduced state space towards lower computational requirements and more interpretable understanding of agent behaviour.

The AYS model has been specifically tuned so that an agent tends towards one of two points:

\begin{align}
    \bullet \; \; \text{Green fixed point} &- \begin{pmatrix} 0 \\ \infty \\ \infty \end{pmatrix}, \quad \quad \quad \quad \bullet \; \; \text{Black fixed point} &- \begin{pmatrix} \frac{\beta}{\theta} \\ \frac{\phi \beta \epsilon}{\theta \tau_A}\\ 0 \end{pmatrix} = \begin{pmatrix} 350 \; GtC \\ 4.84 \times 10^{13} \; \$ yr^{-1} \\ 0 \; GJ \end{pmatrix}
    \label{eqn:fixed_points}
\end{align}

The green fixed point denotes a ``sustainable" future, one where there is no atmospheric carbon but limitless capital and renewable knowledge. The black fixed point however, denotes a stagnant economy solely dependant on fossil fuels. This is a future we ideally want to avoid. Included with these ``drain" points are Planetary Boundaries (PB). The AYS model incorporates one PB set in the reports from \cite{steffen2015planetary} and \cite{rockstrom2009planetary} of a maximum excess atmospheric carbon at $PB_A = 345 \; GtC$, with a social foundation for prosperity from \cite{dearing2014safe} defining a minimum yearly economic output at $PB_Y = 4 \times 10^{13} \; \$ yr^{-1}$ \citep{kittel2021lakes}. For brevity throughout this paper we will make reference to these boundaries as the two PBs, although by definition our economic output boundary is in fact a social goal, not a planetary boundary.
\begin{figure}[h]
  \centering
  \includegraphics[width=0.9\textwidth]{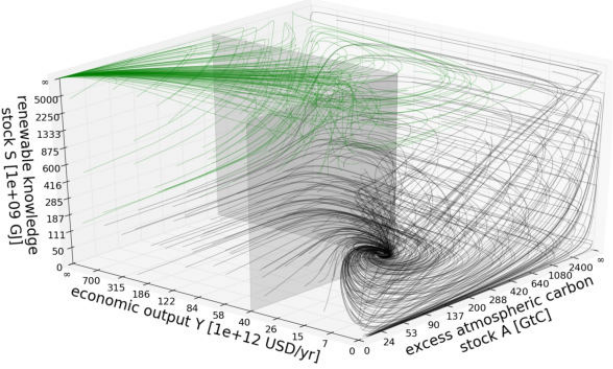}
    \caption[AYS Model Flow Lines]{\small The AYS model state space from \cite{kittel2021lakes}. Translucent grey planes signify the two PBs, and the green and black points denote the fixed point end conditions for a single agent. Whisker lines indicate flow forces within the model, that tend towards either of the two fixed points. The colours showing the flow to the respective fixed points.}
    \label{fig:ays_model}
\end{figure}

 To mimic the current state of the Earth within this model, the starting point is defined as $s_{t=0}=\{240 \; GtC, 7 \times 10^{13} \; \$ yr^{-1}, 5 \times 10^{11} \; GJ\}$. Not only is this starting location very close to the PBs creating a challenging control problem, but also from this location the agent will tend towards the black fixed point if no actions are taken. 
 Figure \ref{fig:ays_model} highlights the AYS environment with black and green fixed points, and the two translucent grey planes indicating the two PBs. 
 \cite{strnad2019deep} and \cite{wolf2022climate} incorporate noise into the starting position over episodes to improve training, however, noise is omitted from the S state variable as this dramatically reduces the agents' ability to learn. \cite{kittel2021lakes} and subsequent work normalised the environment between $0$ and $1$ to prevent numerical explosions. 


We carry this through, normalising the states and then incorporating noise, setting the starting state as: 

\begin{equation}
\label{eqn:start_point}
    s_{t=0} = \begin{pmatrix} 0.5 + \mathcal{U}(-0.05,0.05) \\ 0.5 + \mathcal{U}(-0.05,0.05) \\ 0.5 \end{pmatrix}
\end{equation}

where $\mathcal{U}$ is the uniform distribution.

At its current state the model will tend towards the black fixed point. To avoid this an agent is able to undertake four actions, described in \cite{kittel2021lakes}:

\begin{enumerate}[start=0]
\label{list:actions}
    \item Default - Default parameters are used and the agent follows the flow lines without any resistance.
    \item Degrowth - Economic growth parameter $\beta$ is halved, fluctuating between 3\% and 1.5\% growth.
    \item Energy Transition - Break-even renewable knowledge $\sigma$ is reduced by 31.3\%, equivalent to halving the renewable to fossil fuel energy cost ratio.
    \item Both - The two non default actions are combined within one timestep.
\end{enumerate}

For each integration timestep of the environment, an agent is able to select one of these four options, mimicking an action taken every year \citep{kittel2021lakes}. 

\begin{figure}[H]
\centering
\begin{tikzpicture}[thick, main/.style = {draw, rectangle, font=\LARGE, node distance={45mm}}, main_2/.style = {draw, rectangle, font=\LARGE, node distance={30mm}}, secondary/.style={font=\large, node distance={10mm}}, secondary_2/.style={font=\large, node distance={13mm}}, arrow/.style={line width=0.5mm}, dashed_arrow/.style={line width=0.5mm, dashed}]
\usetikzlibrary{arrows.meta}
\def\myarrowhead{Stealth[length=10mm]}
\node[main] (1) {$\dot{Y}_i = \beta_i Y_i - \theta \xi_i A Y_i$}; 
\node[main] (2) [right of=1] {$U_i = \frac{Y_i}{\epsilon_i}$}; 
\node[main] (3) [right of=2] {$\Gamma_i = \frac{1}{1 + (\frac{S_i}{\sigma_i}) ^ {\rho_i}}$}; 
\node[main_2] (4) [above of=1] {$\dot{A} = \frac{\sum_n E_i}{n} - \frac{A}{\tau_A}$}; 
\node[main_2] (5) [above of=3] {$E_i = \frac{\Gamma_i U_i}{\phi_i}$};  
\node[main_2] (6) [below of=1] {$\dot{S_i} = R_i - \frac{S_i}{\tau_{S,i}}$};
\node[main_2] (7) [below of=3] {$R_i = (1 - \Gamma_i) U_i$};
\node[secondary] [below of=1] {Economic Output};
\node[secondary_2] [above of=2] {Energy Demand};
\node[secondary] [above of=3] {\; \; Fossil Fuel Energy Share};
\node[secondary] [above of=4] {Excess Atmospheric Carbon};
\node[secondary] [above of=5] {Emissions};
\node[secondary] [below of=6] {Renewable Knowledge Stock};
\node[secondary] [below of=7] {Renewable Energy Produced};
\draw[dashed_arrow,->] (1.177) arc (0:330:7mm);
\draw[dashed_arrow,->] (4.177) arc (0:330:7mm);
\draw[dashed_arrow,->] (6.177) arc (0:330:7mm);
\draw[arrow, ->] (1) -- (2); 
\draw[arrow, ->] (2) -- (5); 
\draw[arrow, ->] (2) -- (7); 
\draw[arrow, ->] (3) -- (5); 
\draw[arrow, ->] (5) -- (4); 
\draw[dashed_arrow, ->] (4) -- (1); 
\draw[arrow, ->] (3) -- (7);
\draw[arrow, ->] (7) -- (6); 
\draw[dashed_arrow, ->] (6) -- (3); 
\end{tikzpicture}
\caption{Multi-agent AYS interaction cycle (diagram adapted from \cite{kittel2021lakes}). Block arrows are positive interactions, dashed arrows are negative interactions.}
\label{fig:multi_agent_ays_cycle}
\end{figure}
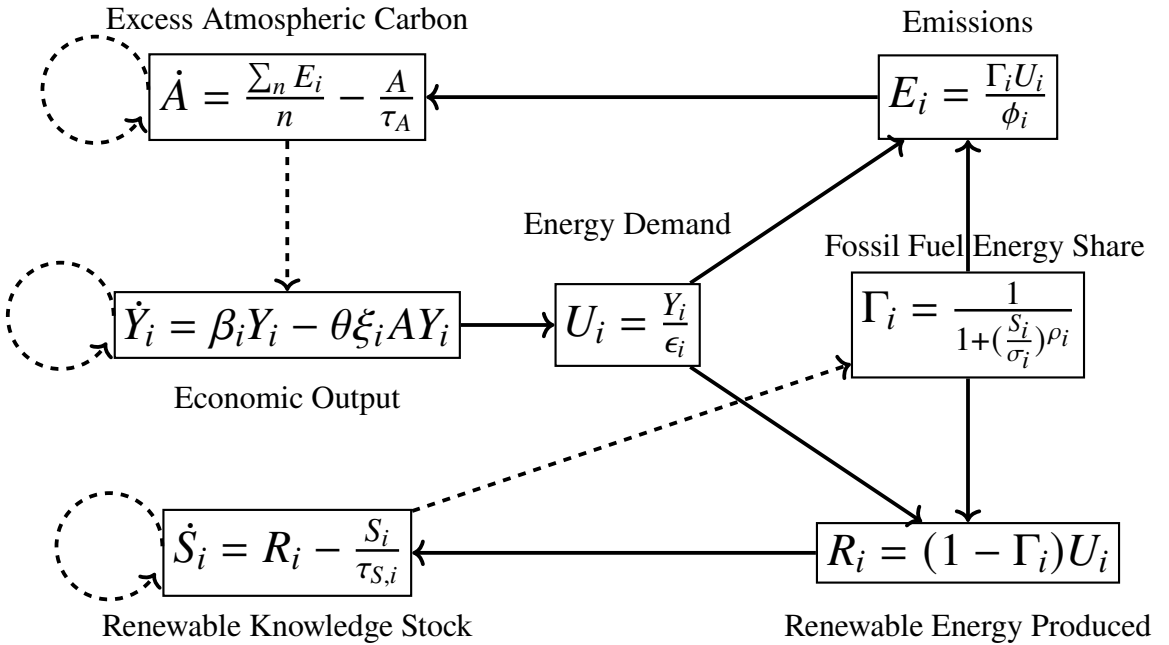

The AYS model in its current format depends on only one agent driving the simulation. We propose an extension enabling simple interactions between multiple agents. Global variables are denoted with no subscript, however, local (to each agent) variables are denoted with a subscript. There is now only one global variable - the excess atmospheric carbon A. Figure \ref{fig:multi_agent_ays_cycle} visualises the extended multi-agent environment differential equation cycle. 

We carry through the same PBs and green fixed point, as they still apply to the global scale. However, the black fixed point is individual to each agent as Equation \ref{eqn:fixed_points} is dependent on individual agent parameters. We have also normalised emissions on the global scale so that we can work within the same parameters as the original AYS model. This is the simplest approach allowing us to focus on interacting with the model rather than heavily editing the model. 
We have adjusted the axes in Figure \ref{fig:ays_model} to enable greater insight when dealing with multiple agents. The S and A axis are swapped and S variable then replaced with Equation \ref{eqn:emissions} for agent dependent emissions $E$. Incorporating emissions visualises the individual impact each agent has towards the shared $A$.

We have adopted the JAX framework \citep{bradbury2018jax}, converting the environment to be fully vectorised, allowing both inference and environment loops to be run on a GPU. The original environment from \cite{kittel2021lakes} utilises an ODE solver to calculate the environment transition at each time step. Due to JAX's default enforcement of single precision floats, there is a discrepancy in the ODE solver results from \cite{kittel2021lakes}, \cite{strnad2019deep}, and \cite{wolf2022climate} as their solver used double precision. However, this precision error has been tested over a wide range of states in the environment, with a minimum value of $0.000$ and maximum of $1.055e^{-05}$. This is a minute discrepancy, so we have assumed parity.

This extended AYS environment can be modelled as a Partially Observable Stochastic Game (POSG) \citep{shapley1953stochastic, hansen2004dynamic}, defined by the tuple $<N, \mathcal S, \mathcal A_1,..., \mathcal A_n, T, R_1,...,R_n, \mathcal O_1,..., \mathcal O_n, \gamma>$, where $N$ is the number of agents, $\mathcal S$ is the set of all possible environmental states, $\mathcal A_1,..., \mathcal A_n$ is the set of possible actions for each agent, $T:\mathcal S \times \mathcal A_1 \times ... \times \mathcal A_n \times \mathcal S \rightarrow \Pi (\mathcal S)$ is the transition distribution, ${R_i}_{i=1}^n$ is the set of reward functions where $R_i: \mathcal{S} \times \mathcal{A} \rightarrow \mathbb{R}$ is the reward function for agent $i$, and $\gamma$ is the discount factor. Each agent $i$ has access to its observation $o^i \in \mathcal O_i$ where $\mathcal O^i$ is the observation set of agent $i$. 

\subsection{MARL Algorithm}

Focusing on DTDE algorithms as stated in the introduction, the Independent Proximal Policy Optimisation (IPPO) algorithm acts as an effective starting point \citep{schulman2017proximal, yu2022surprising}. This relates to $n$ (number of agent) versions of PPO based agents within an environment that do not share parameters between them, so are fully independent. Each ($0$ to $n$) PPO agent \citep{schulman2017proximal} has no awareness of other agents in the system, and since we are in a POSG, only has access to its observations of the environment. 
The state and observation space is a vector of values $\in [0,1]$ relating to the three AYS variables. A is global, but Y and S are independent to each agent leading to the partially observable nature. The action space contains values from the discrete set $\{0, 1, 2, 3\}$ relating to the actions in List \ref{list:actions}. Our previous work in \cite{wolf2022climate} found PPO to achieve impressive results and thus further posits its use within our experiments. Rewards are derived from the "Planetary Boundary" (PB) reward function, maximising the euclidean distance between the agent and the two PBs and a lower bound of 0 on the S parameter. If a boundary is crossed the reward equals 0:
\begin{equation}
    R_{PB} = ||o - o_{PB} || ^2
\end{equation}

where $o$ relates to an individual agent's observations of the environment. As an agent aims to maximise its reward, it looks to achieve a point as far away from the PBs as possible, thus tending towards the green fixed point. Using the PB rather than the limits of the simulation incentivises the agent to avoid the PBs. For further experiments we look at competitive agents and thus need two new reward functions: 

\begin{align}
    R_{maxA} &= o_A \\
    R_{maxY} &= o_Y - PB_Y,
\end{align} 

where $o_A$ is the agent observation of the $A$ variable, $o_Y$ is the agent observation of the $Y$ variable, and $PB_Y$ is the planetary boundary (social goal) for the $Y$ variable. The former directly rewards an agent on the A variable, the excess atmospheric carbon ($GtC$), relating to an entity that prioritises environmental degradation. The latter at maximising the agent's distance to the $Y$ planetary boundary, the economic output ($\text{\$yr}^{-1}$) social goal, which can be seen as an entity that prioritises economic gain over environmental impact.

\subsection{Critical States}

Explainability and interpretability in RL is an open question, with most methods focusing on explaining the neural networks that are used as functional approximators in deep RL \citep{heuillet2021explainability}. There are very few methods that are specific to RL algorithms, and even fewer that are usable rather than purely conceptual \citep{heuillet2021explainability}. 
Critical states, based on \cite{Huang2018}, serves as a form of explainability specific to RL for model policy. This work elaborates that there are a set of few specific states (critical states) in an agent's trajectory in which it greatly matters which action the agent takes \citep{Huang2018}. 
In theory, certain states lead to a large difference between policy outputs over the set of actions. Generally, one action would lead to a much larger policy value than the rest, as the agent is more sure this is the only action option in that state.
We proceed with this method of explainability, as it is crucial to know which locations in a trajectory correspond to the most vital actions for actionable climate policies.
In more concrete terms, the set of critical states $\mathcal C_\pi$ are identified as those with a high \emph{logit difference}, calculated from the outputs of the neural network representation of the agent's policy, mathematically formalised as:

\begin{equation}
    \mathcal C_\pi = \{ s | \max_a \pi_\theta(s, a) - \frac{1}{|\mathcal{A}|}\sum_a\pi_\theta(s, a) > t \}
\end{equation}

where $\pi_\theta(s, a)$ represents the logits of the policy distribution (as output by the actor network), $t$ a critical state threshold, and $\mathcal{A}$ is the set of potential actions.
A requirement is that entropy regularisation is used in the policy objective – without it, policies can collapse prematurely to almost deterministic states, signifying that almost all states are critical \citep{Huang2018}. We have included entropy regularisation into our implementation of PPO, ensuring the policy acts purposefully in critical states and more randomly in others \citep{Huang2018}. 
We expand on the idea of critical states by plotting the logit differences across 1000 sampled trajectories (post-training) to analyse how ``critical" each state is, rather than defining a critical state threshold. The value of this threshold is arbitrary and we prefer to highlight the full range over states, although one could consider states with a logit difference over $0.5$ as the critical states. In particular, we ask: Are there locations in the trajectories that the policy finds more critical than others, and are these critical areas distributed in a way that is interpretable with regard to the agent's behaviour?  
To some extent, this can be loosely interpreted as policy uncertainty, as critical states are those in which the policy has a higher logit difference and is thus more \emph{certain} of the correct action to take. However, we try to avoid using this term, as this method does not provide an exact uncertainty quantification of the policy.

\section{Experimental results}
Our overarching ambition is towards applicable and deployable systems that guide climate policy. Whilst this is an expansive open question that can't be fully answered in this paper, we begin by experimenting on the simplest cases and slowly increase complexity. This lines up the following research questions that we tackle within this work:

\begin{itemize}
    \item RQ1 - Assuming agents are homogeneous (having the same starting state and thus the same initial IAM variables), can they achieve an ``economic and environmental positive future" when acting towards a shared goal through having the same reward functions (a.k.a interacting cooperatively)?
    \item RQ2 - Relaxing agent homogeneity, are cooperative agents still able to achieve a successful future at a similar rate?
    \item RQ3 - Finally, does introducing competition between agents, for example by having reward functions that oppose each other to discourage cooperation, significantly hinder a strategic interaction convergence on reaching the green fixed point?
\end{itemize}

Towards RQ1 our first experiment incorporates increasing numbers of homogeneous cooperative agents into the AYS environment. For RQ2 we repeat the same experiments as RQ1 but allow agents to start in varying locations to each other, initialising an agent's state at different AYS variables, thus mimicking the variability seen between entities/nations in reality. 
Furthering agent heterogeneity we also vary the agent independent values for climate damages $\xi_i$ mimicking agents not all experiencing the same damaging effects as the climate degrades.
Finally for RQ3 we reduce the number of agents in our environment to two to compare varying reward functions and their effects on an agent's ability to reach the green fixed point. Then extend this to three agents highlighting that the trend continues as agent numbers increase. By keeping the number of agents low as well as incorporating the critical states visualisation we show greater insight into the agent's action decisions.

A key theme within our research questions is the ability for an agent to reach the green fixed point. We define the win rate as the percentage of times that the simulation \textit(as a whole) reaches the green fixed point over a set number of episodes. However, the definition of success within this environment is not a Pareto Frontier and instead stakes claims on what is negative or positive, as such we focus in on the environmental positives. For clarity an episode is the collection of timesteps between an initial state and a terminal state, be that due to reaching the green fixed point, breaching a planetary boundary, or reaching the fixed maximum number of steps per episode. We run all experiments for six seeds and plot the average of these seeds with translucent standard error bounds.

\subsection{Experiment 1 - Homogeneous Agents}
We begin by instantiating homogeneous agents, i.e. agents that have the same initial AYS variables. This relates to all agents starting in the same location. Agents here have the same objective towards a common goal, each following the $R_{PB}$ reward function. The greater the distance to the PBs the greater the reward. Agents are not predefined with a top-down restraint that they must cooperate, instead by using a reward with a shared goal we show the emergence of cooperation.

\begin{figure}[h]
  \centering
  \includegraphics[width=0.99\textwidth]{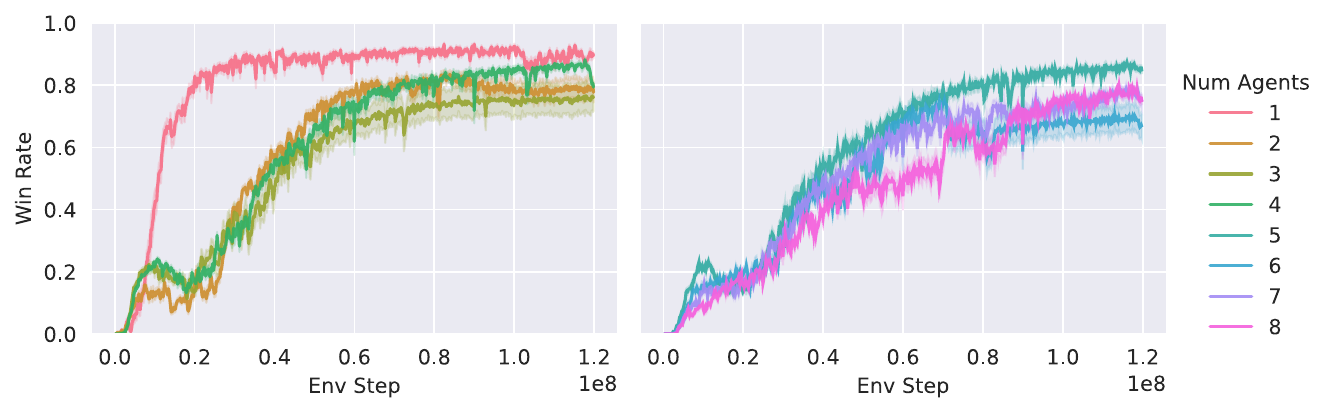}
    \caption[Homogeneous Tests]{\small Homogeneous agent's win rates. Each experiment is run over six seeds with the line corresponding to mean win rate with translucent standard error bounds. Num agents relates to the number of agents in the simulation.}
    \label{fig:homo_tests}
\end{figure}

In Figure \ref{fig:homo_tests} for a single agent case IPPO (which reduces to PPO for one agent) quickly learns a consistent policy, as it avoids any complexity from the non-stationarity of the transition function caused by other agents. Increasing the number of agents (ranging from 2 to 8 agents together), increases training time taken until a consistent policy is reached which can be attributed to the increasing complexity stemming from the non-stationarity and interactions between agents.

\begin{figure}[h]
  \centering
  \includegraphics[width=0.7\textwidth]{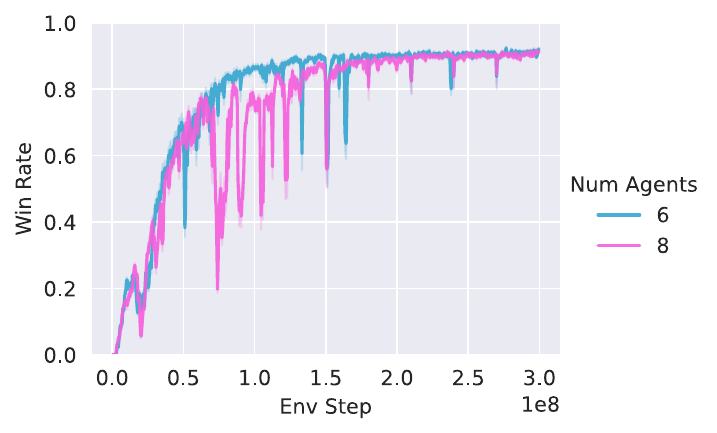}
    \caption[Homogeneous Tests]{\small Homogeneous agent's win rates for a longer range of training steps. These experiments are only run over two seeds due to computational constraints.}
    \label{fig:homo_tests_long_one}
\end{figure}

Figure \ref{fig:homo_tests_long_one} shows (with only two seeds leading to a larger variance during the middle of training) that with enough time steps a similar win rate is achieved between agents. We have not run the experiments in Figure \ref{fig:homo_tests} to a stable state for large numbers of agents due to the computational resources required, and instead focus on a smaller total of agents (and for fewer random seeds) for greater insight. For a singular agent, the win rate after $1.2 \times 10^8$ steps is $87.740\% \pm 8.225$. For six and eight agents after $3 \times 10^8$ steps the win rates are $90.935 \pm 0.010$ and $90.143 \pm 0.035$ respectively. The lower standard deviations here stem from the policy convergence gained from much longer time steps. Answering RQ1 it is clear that agents are able to reach the green fixed point consistently, independently of the number of agents. Cooperation thus emerges between agents, with the shared reward function of a common goal being the only predefined signal towards cooperating.

\subsection{Experiment 2 - Heterogeneous Agents}
Increasing the applicability we now look at heterogeneous, but still cooperative, agents. 
Heterogeneity is very important in the climate domain, especially when dealing with anthropogenic factors as it can apply to: spatial variability, temporal variability, and variability in socio-economic impacts, among others \citep{madani2013modeling}. 
The various sources of heterogeneity between agents in the AYS MARL environment are: AYS variables, AYS parameters, Reward Functions, MARL algorithm.
Varying the AYS variables and parameters can be seen as representing different traits of a representative agent, for example a larger initial $Y$ may indicate an economically wealthy entity. Similarly changing for the economic growth parameter $\beta$ again represents an entity with increased economic function. There are limitless combinations one could make from these for experimentation. Values could also be based on real world data to provide an in silico entity representation, or verify results on a well known case study. 
Reward functions represent what an entity may "value" or be looking to optimise for, changing these between agents can lead to conflicting behaviour as these may directly oppose one another.
Finally we can represent each agent with different MARL algorithms since we are constrained to the use of DTDE algorithms which have no overarching centralised controller. For example we could represent certain agents with less capable algorithms to understand the effect on the resulting equilibrium. 
We do not adjust the MARL algorithm, using PPO for all, as we want to understand some of the limitations of RL specific algorithms being applied to MARL in this domain.
Instead we vary the AYS variables and parameters, with our subsequent experiments adjusting the reward function.
Agents can start at any location within the predefined uniform distribution of starting points. A new starting point is sampled at each episode.

\begin{figure}[h]
  \centering
  \includegraphics[width=0.99\textwidth]{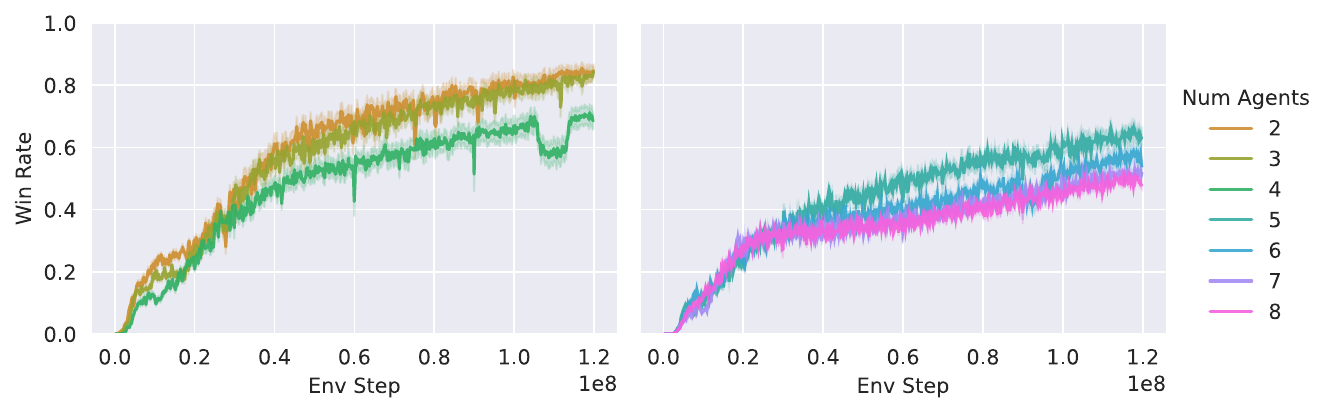}
    \caption[Heterogeneous Tests]{\small Heterogeneous agent's win rates. We have omitted the single agent scenario as these results match between homogeneous and heterogeneous starting points. Each experiment is run over six seeds with the line corresponding to mean win rate with translucent standard error bounds.}
    \label{fig:hetero_tests}
\end{figure}

\begin{figure}[h]
  \centering
  \includegraphics[width=0.7\textwidth]{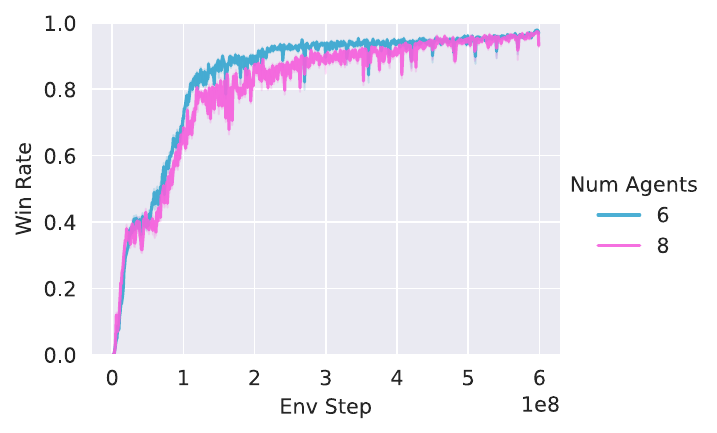}
    \caption[Homogeneous Tests]{\small Heterogeneous agent's win rates for a longer range of training steps. These experiments are only run over two seeds due to computational constraints.}
    \label{fig:hetero_tests_long_one}
\end{figure}

Figure \ref{fig:hetero_tests} shows that scaling up agents here has a larger impact on the win rate due to the more complex heterogeneous nature of the agents. Still again with enough timesteps agents reach a consistent policy, as seen in Figure \ref{fig:hetero_tests_long_one}. Win rates for six and eight agents after $6 \times 10^8$ steps are $93.007 \pm 0.054$ and $94.121 \pm 0.067$ respectively. Closely matching the results found in Experiment 1.

Multiple heterogeneous agents acting towards the same goal have similar performance to a singular agent, although require a much longer set of episodes for convergence due to the increased complexity.
Here we prove that RQ2 is possible, without any loss of performance.

Furthering these experiments we also look at heterogeneity in the AYS parameters, specifically scaling the agent independent climate damage $\xi_i$. We carry over the same heterogeneous starting point variation as in the previous experiment and only focus on two agents together. 
In reality negative environmental effects such as extreme weather scenarios or rising water levels that impact economic output may affect certain regions more than others \citep{dellink2019sectoral}.
In the worst scenarios the biggest polluters may rarely see the negative climate effects, which are instead fully experienced at other geographical locations.
To naively model this we scale the climate damage parameter $\xi_i$ between $0$ and $1$, the former an extreme case where the economy is not affected by parameter $A$, and the latter the usual AYS ODE dynamics.

\begin{figure}[h]
  \centering
  \includegraphics[width=0.99\textwidth]{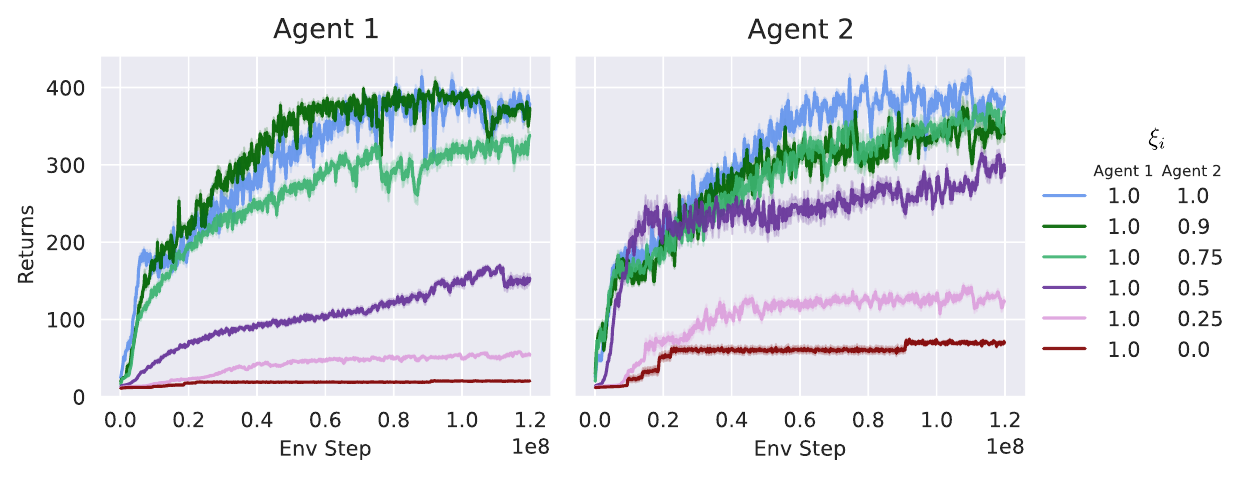}
    \caption[Homogeneous Tests]{\small Returns for each agent for the climate damages parameter $\xi_i$ experiments. Agent $1$ episode returns are on the left, which always has $\xi_1 = 0$. Agent $2$ episode returns are on the right where $\xi_2$ varies between $0$ and $1$ as per the figure legend. }
    \label{fig:climate_damage_tests_returns}
\end{figure}

\begin{figure}[h]
  \centering
  \includegraphics[width=0.7\textwidth]{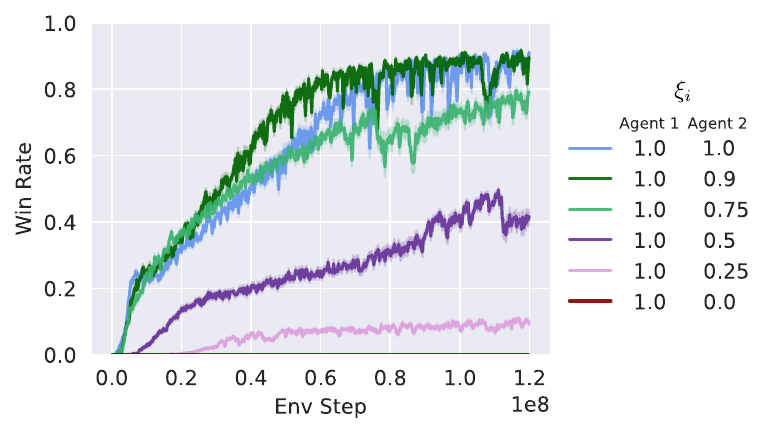}
    \caption[Homogeneous Tests]{\small Overall win rates for a two agent scenario in which both agents follow the $R_{PB}$ reward function, but have different climate damage parameters $\xi_i$ for each experiment. Six combinations of $\xi_i$ are tested.}
    \label{fig:climate_damage_tests_win_rate}
\end{figure}

\begin{figure}[h]
\centering
\begin{subfigure}[t]{0.495\textwidth}
  \centering
  \includegraphics[width=\linewidth]{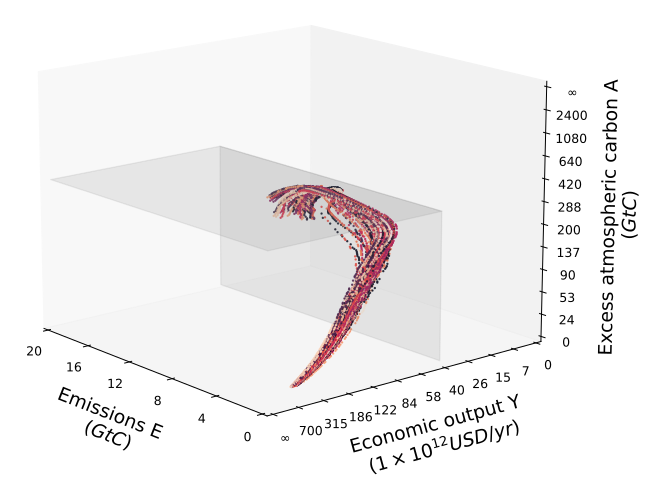}
  \caption{Experiment 1, Agent 1 with $\xi_i = 1$}
\end{subfigure}%
\hfill
\begin{subfigure}[t]{0.495\textwidth}
  \centering
  \includegraphics[width=\linewidth]{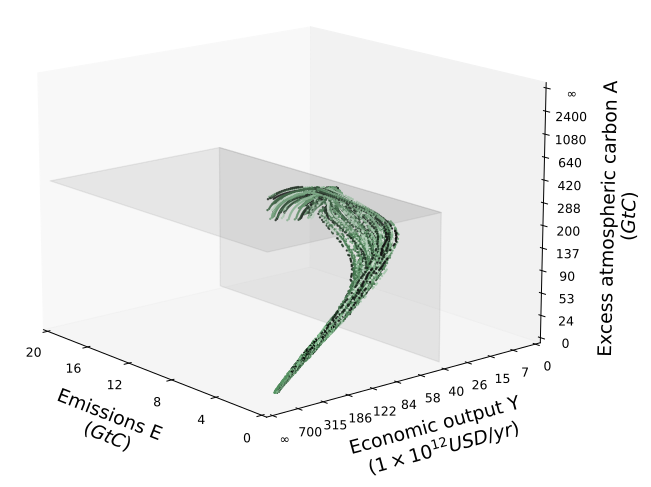}
    \caption{Experiment 1, Agent 2 with $\xi_i = 1$}
\end{subfigure}
\medskip  
\begin{subfigure}[t]{0.495\textwidth}
  \centering
  \includegraphics[width=\linewidth]{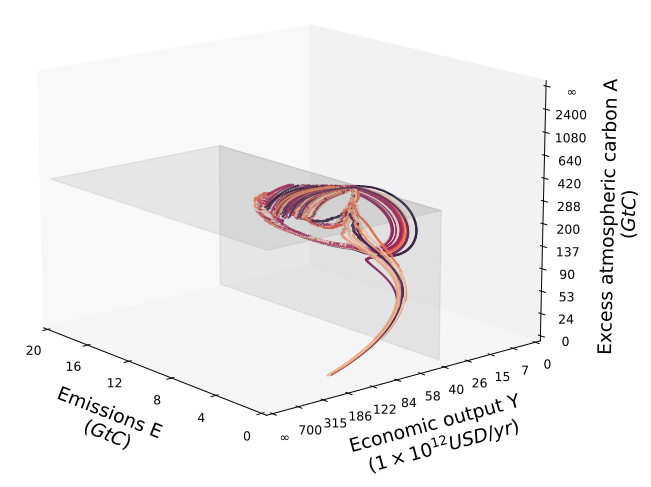}
    \caption{Experiment 2, Agent 1 with $\xi_i = 1$}
\end{subfigure}%
\hfill
\begin{subfigure}[t]{0.495\textwidth}
  \centering
  \includegraphics[width=\linewidth]{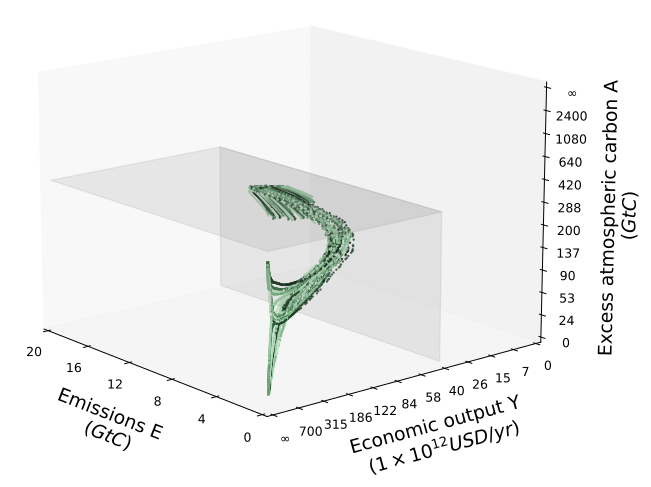}
    \caption{Experiment 2, Agent 2 with $\xi_i = 0.25$}
    \label{fig:climate_trajectories_almost_maxy}
\end{subfigure}
\caption{\small Trajectory plots for two cooperative agents, both following the $R_{PB}$ reward function. Agent $1$ has red trajectories, and Agent $2$ has green. The variation in colour for each agent signifies trajectories from different episodes. We have visualised a sample of $1000$ episodes (trajectories) to indicate the distribution of trajectories. The grid row relates to experiments that contain both agents together. In the upper row both the agents experience the same climate damages, with $\xi_i = 1$ for each. In the lower row Agent $1$ has $\xi_1 = 1$ and Agent $2$ has $\xi_2 = 0.25$. 
The green fixed point is situated on the lowest vertex of the Figures, where $E = 0$, $Y = \infty$, and $A = 0$.
The distribution of starting states is near the middle of the Figures, where $E \approx 10$, $Y \approx 60$, and $A \approx 250$.}
\label{fig:climate_damage_pathways}
\end{figure}

Figure \ref{fig:climate_damage_tests_returns} indicate that as an agent is impacted less by climate damages, i.e. as $\xi_i$ tends towards 0, it gains more independent return (total individual reward over an episode) than the other agent that has $\xi_i = 1$.
Importantly though it comes at the cost of globally reaching the green fixed point, even with cooperative reward functions, as seen in Figure \ref{fig:climate_damage_tests_win_rate}.
As $\xi_i$ reduces in the AYS ODE interaction Figure \ref{fig:multi_agent_ays_cycle}, $Y$ becomes less affected by the value of $A$ which has knock on effects in further increasing an agent's own Emissions $E$.
However an agent therefore also receives less signal in the observations about how the $A$ variable affects the $Y$ variable, and how this all relates to its own actions and reward function.
Therefore these agents seem to prefer maximising $Y$ as they are unaware of the impact this has on $A$.
In Figure \ref{fig:climate_damage_pathways} one can see how the trajectories evolve from a two agent scenario both following $R_{PB}$ and having $\xi_i$ of $1$, to very different pathways when $\xi_2$ is $0.25$ for Agent $2$.
Interestingly the trajectories for Agent $2$ in Figure \ref{fig:climate_trajectories_almost_maxy} are very similar to those of an agent following the $R_{maxY}$ reward function, with example trajectories found in Figure \ref{fig:pbmaxy_agent_2_pb_actions} and \ref{fig:pbmaxy_agent_2_pb_critical_states}, even though the agent is still following $R_{PB}$.
Without staking too many claims in reality, an agent that has minimal understanding of how the actions it takes impact the environmental variable on a global scale, will be unable to enact the desired actions to reach the ``climate positive" future.

\subsection{Experiment 3 - Competitive Agents}
We have shown that agents are able to consistently reach the green fixed point when working together. However, how will they fare when dealing with more competitive agents, e.g. ones that prioritise capital over detrimental environmental effects? Or in an extreme (yet slightly unrealistic) case, agents that only care to maximise the excess carbon in the atmosphere. For this, we use the two other reward functions: $R_{maxY}$ and $R_{maxA}$.
The former rewarding an agent for maximising the distance to the $Y$ planetary boundary, the economic output ($\text{\$yr}^{-1}$) social goal.
The latter rewarding an agent for maximising the $A$ variable, the excess atmospheric carbon ($GtC$). 
We also assume that agents start in heterogeneous locations as our experiments have shown this does not negatively impact the win rate. 
The choice of $R_{maxA}$ may be a peculiar one, but we have included the experiments to show more adversarial behaviour than can be expected with $R_{maxY}$. The definition of $R_{PB}$ in some ways includes maximising $Y$, or at least ensuring that the agent avoids the $Y$ social goal boundary, and as such $R_{maxY}$ can be seen as a mixed motivation reward function. Whereas $R_{maxA}$ greatly opposes the aims of $R_{PB}$, leaning towards more competition. This choice helps us understand the performance of the IPPO algorithm in these more challenging competitive scenarios, which will arise in future applications.

\begin{figure}[h]
  \centering
  \includegraphics[width=0.7\textwidth]{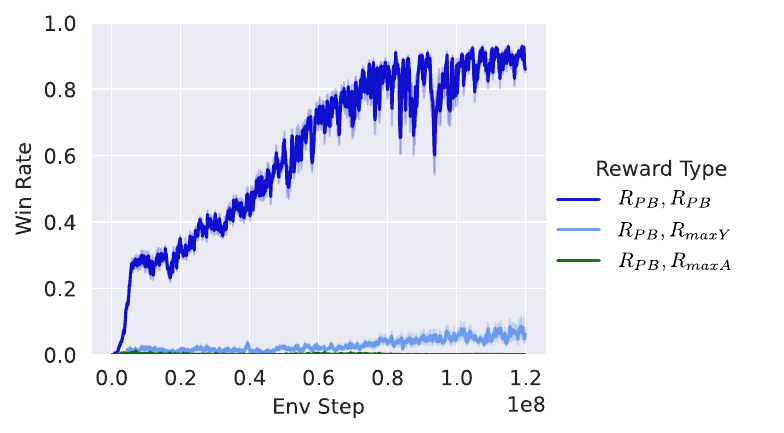}
    \caption[Competitive Two Agents]{\small Experiments combining reward types for a two agent scenario, the first agent always follows the $R_{PB}$ reward function. Each run has two agents relating to the respectively labelled reward type.}
    \label{fig:comp_2_agents}
\end{figure}

As seen in previous experiments and in Figure \ref{fig:comp_2_agents}, two agents following $R_{PB}$ consistently reach the green fixed point. Interestingly agents following $R_{maxY}$ are also able to reach the green fixed point, although at a much reduced capacity. This is due to the AYS environment, wherein the $Y$ variable is directly driven by the atmospheric carbon $A$, greatly incentivising an agent to reduce $A$ in order to maximise $Y$. 

However, as we would unfortunately expect, an agent that only aims to maximise its carbon output (following $R_{maxA}$) overrules any potential climate positive actions from the $R_{PB}$ following agent. This clearly highlights the need for cooperation, or at the least, ways to shape "opponents" actions to more closely align to the desired behaviour.

\begin{figure}[h]
  \centering
  \includegraphics[width=0.7\textwidth]{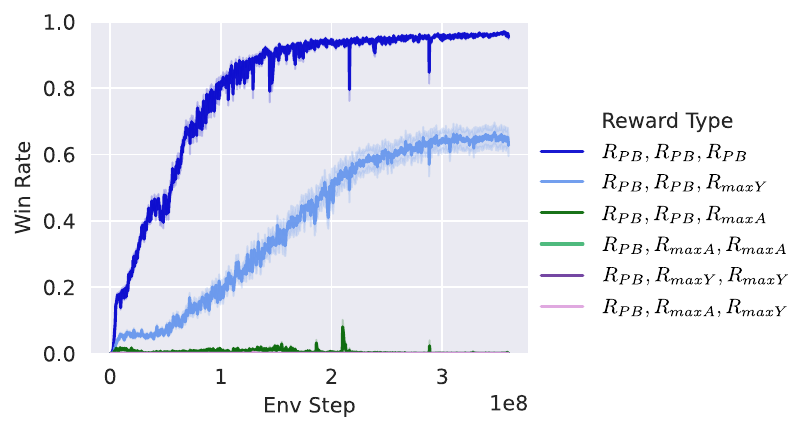}
    \caption[Competitive Three Agents]{\small Experiments combining reward types for a three agent scenario, the first agent always follows the $R_{PB}$ reward function. Each run has three agents relating to the respectively labelled reward type. }
    \label{fig:comp_3_agents}
\end{figure}

In Figure \ref{fig:comp_3_agents} a similar trend carries over with an increasing number of agents. Agents that work together on a shared goal succeed but agents that have different incentives fail, although combinations of a majority of $R_{PB}$ with $R_{maxY}$ have the potential to succeed but at a much reduced rate. Our results confirm RQ3 - increasing competition reduces the ability for agents to reach the green fixed point. Highlighting the need for the use of algorithms with increased opponent awareness over IPPO to improve performance.

In RL defining the reward can be tricky, as agents can "hack" these values and act in non-predictable ways \citep{skalse2022defining, laidlaw2024preventing}. Due to the possibility for early termination from reaching goal states or boundary conditions before the max number of time steps, if agents aren't correctly given potential future rewards they can be incentivised to take "longer" in the environment as there are no temporal negatives. This was clear in some competitive environments where without the notion of discounted future rewards, agents following the $R_{PB}$ would receive more reward if they never reached the green fixed point but slowed down the impact of an agent following $R_{maxA}$. 
Therefore we use discounted rewards within this environment.
Correctly defining rewards is relatively easy here but a key question for future applications is how to quantify rewards.


\subsection{Experiment 4 - Critical States}

\begin{figure}[h]
\centering
\begin{subfigure}[t]{0.495\textwidth}
  \centering
  \includegraphics[width=\linewidth]{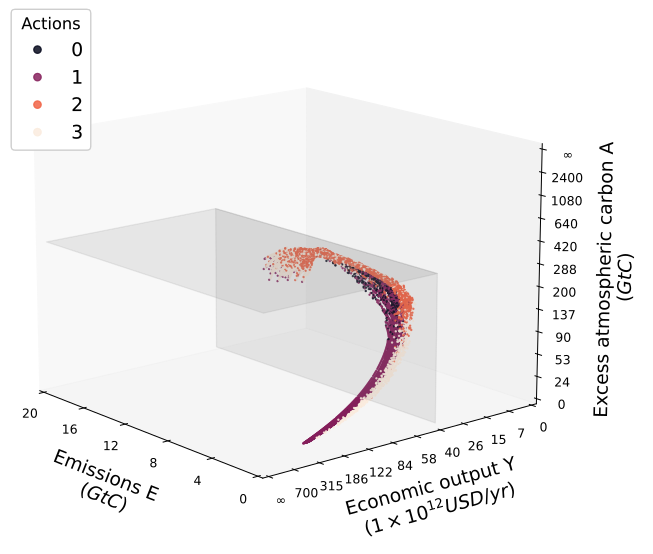}
  \caption{Agent 1 following $R_{PB}$}
\end{subfigure}%
\hfill
\begin{subfigure}[t]{0.495\textwidth}
  \centering
  \includegraphics[width=\linewidth]{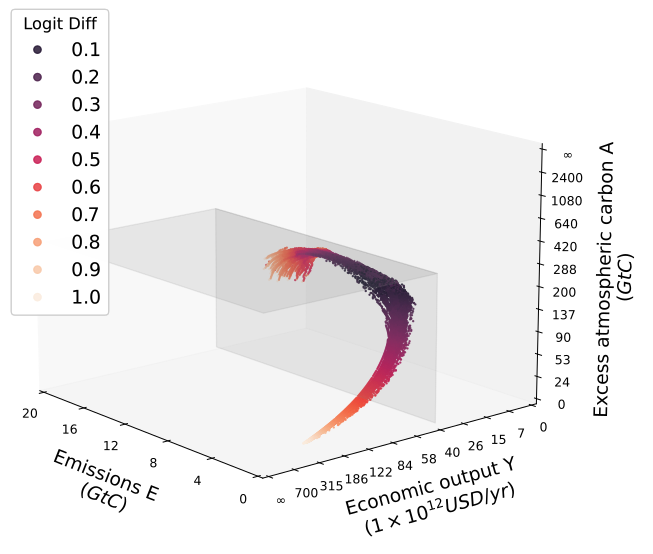}
    \caption{Agent 1 following $R_{PB}$}
    \label{fig:pbpb_agent_1_pb_critical_states}
\end{subfigure}
\medskip  
\begin{subfigure}[t]{0.495\textwidth}
  \centering
  \includegraphics[width=\linewidth]{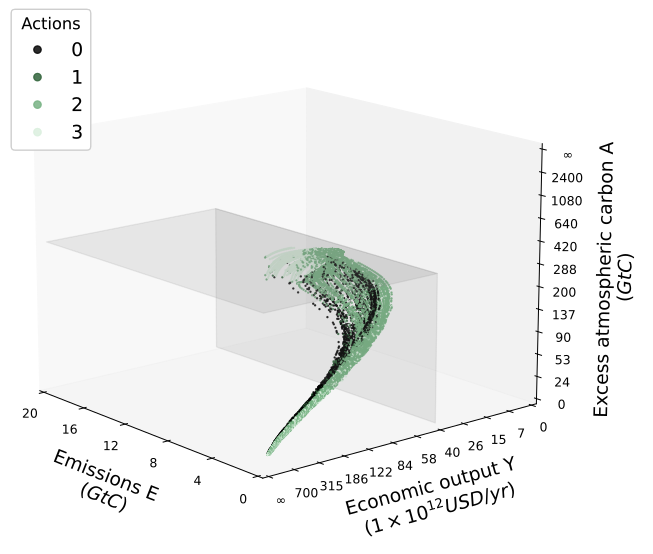}
    \caption{Agent 2 following $R_{PB}$}
    \label{fig:pbpb_agent_2_pb_actions}
\end{subfigure}%
\hfill
\begin{subfigure}[t]{0.495\textwidth}
  \centering
  \includegraphics[width=\linewidth]{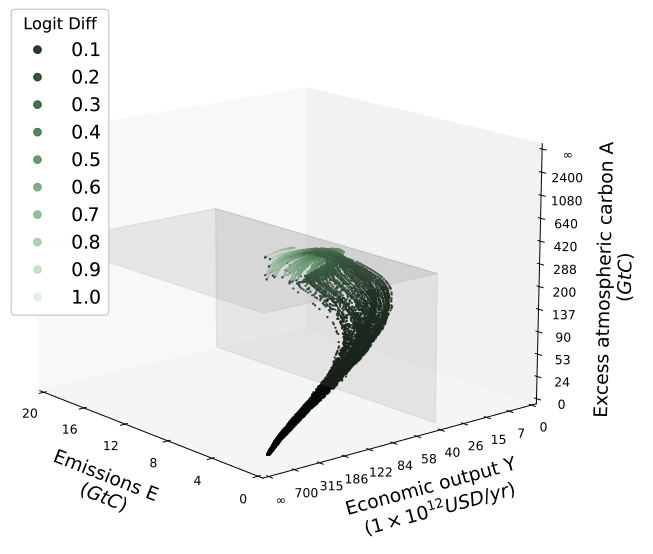}
    \caption{Agent 2 following $R_{PB}$}
    \label{fig:pbpb_agent_2_pb_critical_states}
\end{subfigure}
\caption{\small Critical state plots for two cooperative agents, both following the $R_{PB}$ reward function. Figures on the left hand side represent the actions taken at certain points along the trajectory. Reference List \ref{list:actions} that details all potential actions. Figures on the right hand side indicate scales of logit difference in the agent's policy action distribution, defined as the Logit Diff. Darker colours relate to lower logit difference, with the colour gradation normalised over agents.}
\label{fig:two_agents_pb_pb}
\end{figure}

\begin{figure}[h]
\centering
\begin{subfigure}[t]{0.495\textwidth}
  \centering
  \includegraphics[width=\linewidth]{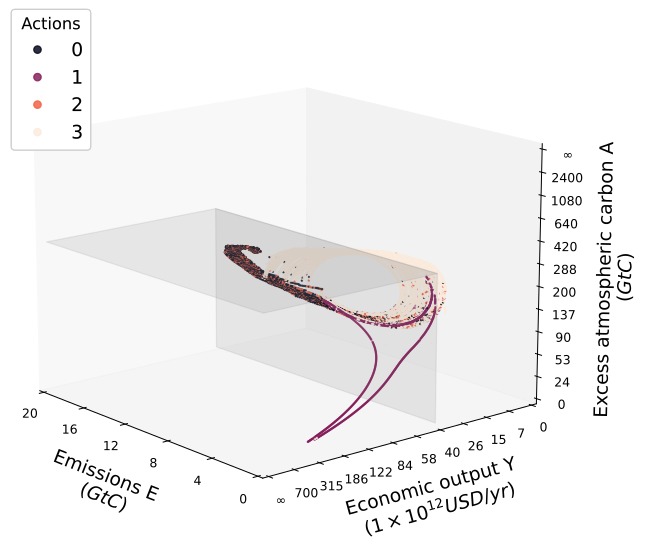}
  \caption{Agent 1 following $R_{PB}$}
\end{subfigure}%
\hfill
\begin{subfigure}[t]{0.495\textwidth}
  \centering
  \includegraphics[width=\linewidth]{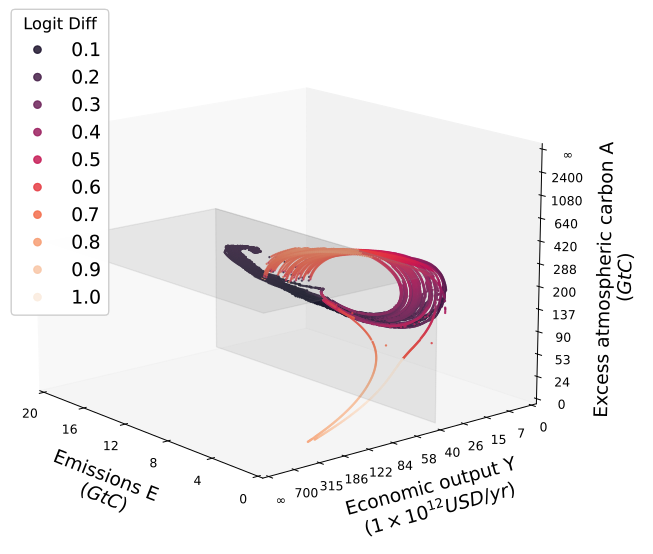}
  \caption{Agent 1 following $R_{PB}$}
\end{subfigure}
\medskip  
\begin{subfigure}[t]{0.495\textwidth}
  \centering
  \includegraphics[width=\linewidth]{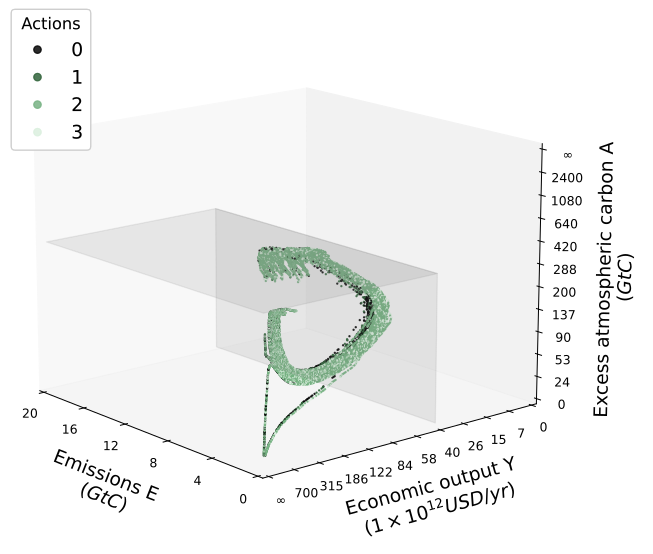}
  \caption{Agent 2 following $R_{maxY}$}
  \label{fig:pbmaxy_agent_2_pb_actions}
\end{subfigure}%
\hfill
\begin{subfigure}[t]{0.495\textwidth}
  \centering
  \includegraphics[width=\linewidth]{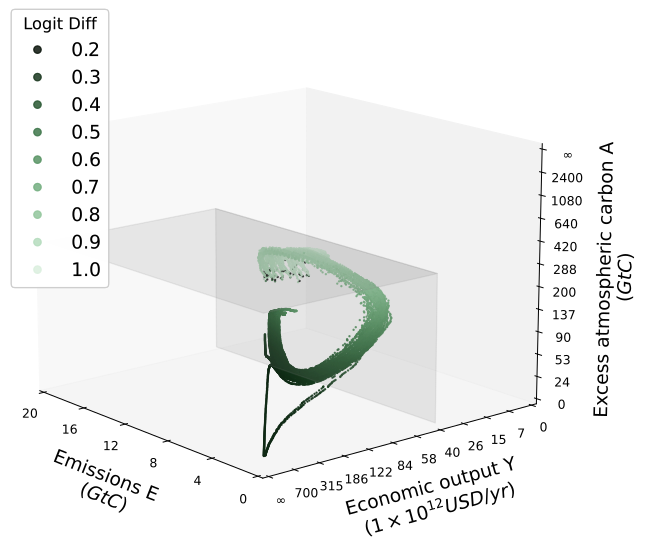}
  \caption{Agent 2 following $R_{maxY}$}
  \label{fig:pbmaxy_agent_2_pb_critical_states}
\end{subfigure}
\caption{Critical states for two competitive agents, where the agents follow the $R_{PB}$ and $R_{maxY}$ reward functions respectively.}
\label{fig:two_agents_pb_maxy}
\end{figure}

Finally, we look into interpreting the behaviour of the agents and attempting to understand failure points. 
To this end, we visualise how ``critical" states are along a sample of trajectories of trained agents in Figures \ref{fig:two_agents_pb_pb} and \ref{fig:two_agents_pb_maxy}. 
Images on the left column represent actions taken at certain points in the trajectory, with images on the right column highlighting the logit difference over actions of the agent's policy.
Darker colours relate to areas in which the policy has a lower logit difference, with increasing difference as the colour lightens. 
The colour gradient scale is normalised over agents. Agents are separated over rows in the multi-grid figure each with their own respective colour map, and the agent's reward function is set as the figure caption. 
To enable a margin of tolerance for reaching the green fixed point, it is defined in the simulation as a ball instead of a singular point. In each critical states figure, the number of displayed agents correlates with the number of agents that were in the simulation – we have not, for example, sampled two agents from a ten agent simulation.

To evaluate these trajectory plots and the quality of explanations that they produce, we establish a set of evaluation metrics consisting of explanation \emph{consistency} and \emph{fidelity}, adapted from \cite{Islam2020} and defined as follows:

\begin{itemize}
    \item Consistency: How consistent are the plots (explanations) between the agents in an experiment?
    \item Fidelity: Are the plots (explanations) logically aligned with the behaviour of the agents?
\end{itemize}

In the context of our experiments, we assess \emph{consistency} between two heterogeneous agents in cooperative and competitive settings and – we note that the same can be done for homogeneous agents as well. The metric \emph{fidelity} more specifically refers to whether the plots accurately represent the nature of the attributes contributing to agent behaviour, such as reward type and location in the trajectory (and accordingly prior knowledge).

With the two agent experiments, it is clear that when agents cooperate (i.e. both follow $R_{PB}$), the simulation as a whole consistently reaches the green fixed point, although different trajectories are able to also succeed. For agent 1, as seen in Figure \ref{fig:pbpb_agent_1_pb_critical_states}, it is clear there is a high logit difference at the start and end of the simulation, signifying the most critical states in which the agent constantly makes the same action. The lowest occurs during the middle phase as the agent passes close to the economic planetary boundary. On the other hand, \ref{fig:pbpb_agent_2_pb_critical_states} shows an agent with the same reward function having similar difference at the beginning but with much lower logit difference towards the end, even though it still takes a consistent action as seen in Figure \ref{fig:pbpb_agent_2_pb_actions}. This emphasises the importance of pairing the consistent action taken with the logit difference for each timestep. 

This indicates a relatively high level of explanation consistency, as the logit difference for both agents are similar until they start to reach the green fixed point – as such, they also have critical states at similar points in their respective trajectories. With regard to explanation fidelity, it is also logical that both agents would be experiencing areas of critical states near the start (corresponding with the action that takes both non-default actions) and then move to lower logit difference levels, as without prior knowledge, the immediate ideal action of the $R_{PB}$ agent is to move away from the planetary boundaries.

For competitive agents, we focus on the $R_{PB}$ and $R_{maxY}$ two agent experiments in Figure \ref{fig:two_agents_pb_maxy} since they show the greatest insight.
Performance is much worse, with only one or two trajectories reaching the green fixed point. This matches the results found in Figure \ref{fig:comp_2_agents} that show a win rate of $~7\%$, similarly matching the ratio of successful trajectories in Figure \ref{fig:two_agents_pb_maxy}. However, it is clear that the agent following $R_{maxY}$ consistently chooses the Energy Transition action so it can maximise its reward. On the other hand, the agent following $R_{PB}$ is unable to have enough effect on the other agent and the environment to reach the green fixed point. On the rare occasions that it does reach the green fixed point, it is confident in its action selection. 

This experiment resulted in high explanation consistency as well, with both agents experiencing similar logit difference levels throughout their trajectories. The exception to this occurs in the few trajectories that reach the green fixed point, where the $R_{PB}$ agent experiences much higher logit difference than the $R_{maxY}$ agent. In terms of explanation fidelity between the actions taken and the logit differences, this also makes sense – while the $R_{PB}$ agent learns all of the environmental attributes, the $R_{maxY}$ agent is focused on maximising the distance from the economic output planetary boundary.

\section{Discussion}
It is clear when constraining agents to have the same objective working towards a common ``climate positive" goal, the green fixed point is consistently reached. This is a promising result but does not carry over once competition is introduced. 
From visualising the critical states figures, agents have lower logit difference when dealing with other agents with differing reward functions, but also have a similar trend even when dealing with others cooperating. 
Combining this insight with the fact we are using IPPO, agents have no explicit understanding of the other agents in the environment. Within basic DTDE methods (like IPPO) other agents are modelled as part of the environment and without an understanding of the consequences of their policies, their actions exacerbate the stochasticity of the environment in the observations of the ego agent. For Centralised Training Decentralised Execution (CTDE) algorithms, there exists a centralised policy between agents during training that reduces the non-stationarity in the transition distribution.
Tackling non-stationary in DTDE algorithms is an open question, with a few types of well researched approaches \citep{papoudakis2019dealing}.
One of which being opponent modelling \citep{albrecht2018autonomous}, where approximate policies are learnt of other agents through historical data and can be used to reduce the effect of non-stationarity, dependent on the validity of the opponent models.
However these can often be sample inefficient and do not explicitly guide exploration to gain an improved understanding of the other agent's desires.
Another branch of MARL research looks into opponent shaping \citep{lu2022model}, how can an ego agent \emph{shape} the behaviour of other agents, through its own actions, to more closely align with its goals.
This approach would have great weight in this domain, as an agent can attempt to steer all agents in the IAM environment towards a ``climate positive future" even with reward functions that may directly oppose this trajectory.

More intricate algorithms however raise issues due to scaling, a primary issue with MARL due to the exponential growth of agent interactions \citep{christianos2021scaling}. 
There is generally an inverse relationship between algorithm capability (e.g. opponent awareness or more principled exploration) and scalability.
Similarly as the IAM complexity increases, most certainly will the MARL state and action spaces which also hinder scalability.
This is a large open question in MARL with many techniques focusing on graph based approaches to balance local and global interactions \citep{nayak2023scalable, ma2024efficient}.
In the application to IAMs we could also take different viewpoints.
One looks at highly abstracted global level IAMs e.g. continents/countries on a world model. We therefore have smaller agent numbers and can focus on more capable algorithms for the more complex global IAMs. Compute more easily covers the large state and action spaces required for complex environments as numbers of agents (and agent interactions) are lower. We mention in the introduction how this could be expanded by imitation learning representative world states from historic data to train against.
Another viewpoint looks at larger numbers of agents (e.g. in the thousands and more) with local scale IAMs, but at the cost (at this current stage) of agent algorithm capability for scalability. Although there is extensive work in this vein such as in multi-agent driving simulations \citep{kazemkhani2024gpudrive} and massively multiplayer online games \citep{suarez2019neural}.
With current work in creating a Digital Twin of Earth \citep{bauer2021digital} that aims to incorporate a wide range of in silico human activity it is clear that scalable agents are needed. 

As these simulations can be used for evidence-based policy, ensuring their validity is important, but how do we assess their uncertainty? Comparing critical states between similar reward functions shows the variability even between agents that appear to follow similar trajectory planning within the set environment, highlighting the poor representation of the policies uncertainty.
The concept of explainability itself has been heavily debated in literature – some believe that rather than attempting to explain black-box models, we should instead just use more intrinsically explainable and transparent models, as explanations can be inconsistent or misleading (\cite{Rudin2019}). In the context of arguments resembling this one, the pitfalls of explainability methods largely fall on post-hoc methods. Potential drawbacks with post-hoc explanations include explanations that are inconsistent based on the method used to generate them, as well as explanations that do not make sense to humans (\cite{Li2018}).

In addition, most post-hoc explainability methods do not provide a fully explainable picture of the model – with the critical states experiment that we performed in this paper, the plots resemble ‘summary statistic’-like results that we can interpret and use to generate explanations for model policy (\cite{Rudin2019}). But we question whether this truly enhances the explainability of a model and correctly quantifies the uncertainty, prompting the question of whether we can deem these explanations to be accurate when they fail to encompass the entire model. While there is potential for the application of these explainability methods, further work is required here, such as exploring more intrinsically explainable methods.



\section{Conclusion}
This paper presents a step towards creating actionable and deployable systems to guide climate policy. Extending on previous work that focused on a single agent scenario we have found that within the bounds of cooperation, and the confines of this environment, multiple agents are consistently able to reach a ``climate positive" future. This ability to craft policy trajectories may help inform policy makers of potential outcomes of prospective plans, with explicit results that can be used as evidence. As is key with any technology used for policy, failure modes and uncertainty must be quantified so that results can be used. To this end, we applied the critical states experiments to gain insight into the policy of the RL model. However, there are strong limitations of this current MARL and interpretability approach and as such we posited various future directions that must be researched if we are to use this technology to guide real policy. 
A key issue with either MARL, ABM, or Optimal Control explored IAMs are scalability, an inherent challenge with MARL itself. Whilst we have no concrete answer to this question, we guide our future work in exploring scalable techniques that still ensure deep exploration of inter-agent behaviour.
However, focusing on global scale low agent number IAMs, this technology could currently be used with data driven stylised world regions to forecast potential policy or action pathways towards a desired outcome.
We hope this is a promising start towards the use of algorithms to support politically guiding the earth's trajectory onto a habitable and stable future.

\begin{Backmatter}


\paragraph{Funding Statement}
James Rudd-Jones is supported by grants from the UK EPSRC-DTP (Award 180330).

\paragraph{Competing Interests}
The authors declare none.

\paragraph{Data Availability Statement}
Our code is publicly available on github at \protect\url{https://github.com/JamesR-J/multi_agent_climate_pathways}


\paragraph{Author Contributions}
Conceptualisation: J.R-J., F.T., M.P-O.; Formal analysis: J.R-J., F.T., M.P-O.; Investigation: J.R-J., F.T., M.P-O.; Methodology: J.R-J., F.T., M.P-O.; Software: J.R-J., F.T.; Supervision: M.P-O; Validation: J.R-J.; Visualisation: J.R-J., Writing–original draft: J.R-J., F.T.; Writing–review and editing: J.R-J., F.T., M.P-O..

\printbibliography

@article{dowlatabadi1995integrated,
  title={Integrated assessment models of climate change: An incomplete overview},
  author={Dowlatabadi, Hadi},
  journal={Energy Policy},
  volume={23},
  number={4-5},
  pages={289--296},
  year={1995},
  publisher={Elsevier}
}

@article{zhang2022ai,
  title={AI for Global Climate Cooperation: Modeling Global Climate Negotiations, Agreements, and Long-Term Cooperation in RICE-N},
  author={Zhang, Tianyu and Williams, Andrew and Phade, Soham and Srinivasa, Sunil and Zhang, Yang and Gupta, Prateek and Bengio, Yoshua and Zheng, Stephan},
  journal={arXiv preprint arXiv:2208.07004},
  year={2022}
}

@article{strnad2019deep,
  title={Deep reinforcement learning in World-Earth system models to discover sustainable management strategies},
  author={Strnad, Felix M and Barfuss, Wolfram and Donges, Jonathan F and Heitzig, Jobst},
  journal={Chaos: An Interdisciplinary Journal of Nonlinear Science},
  volume={29},
  number={12},
  pages={123122},
  year={2019},
  publisher={AIP Publishing LLC}
}

@article{wolf2022climate,
  title={Can Reinforcement Learning support policy makers? A preliminary study with Integrated Assessment Models},
  author={Wolf, Theodore and Nardelli, Nantas and Shawe-Taylor, John and Perez-Ortiz, Maria},
  journal={arXiv preprint arXiv:2312.06527},
  year={2023}
}

@article{madani2013modeling,
  title={Modeling international climate change negotiations more responsibly: Can highly simplified game theory models provide reliable policy insights?},
  author={Madani, Kaveh},
  journal={Ecological Economics},
  volume={90},
  pages={68--76},
  year={2013},
  publisher={Elsevier}
}

@article{rockstrom2009planetary,
  title={Planetary boundaries: exploring the safe operating space for humanity},
  author={Rockstr{\"o}m, Johan and Steffen, Will and Noone, Kevin and Persson, {\AA}sa and Chapin III, F Stuart and Lambin, Eric and Lenton, Timothy M and Scheffer, Marten and Folke, Carl and Schellnhuber, Hans Joachim and others},
  journal={Ecology and society},
  volume={14},
  number={2},
  year={2009},
  publisher={JSTOR}
}

@article{steffen2015planetary,
  title={Planetary boundaries: Guiding human development on a changing planet},
  author={Steffen, Will and Richardson, Katherine and Rockstr{\"o}m, Johan and Cornell, Sarah E and Fetzer, Ingo and Bennett, Elena M and Biggs, Reinette and Carpenter, Stephen R and De Vries, Wim and De Wit, Cynthia A and others},
  journal={Science},
  volume={347},
  number={6223},
  pages={1259855},
  year={2015},
  publisher={American Association for the Advancement of Science}
}

@misc{Theevide7:online,
author = {Luz, Sigourney},
title = {The evidence is clear: the time for action is now. We can halve emissions by 2030. — IPCC},
howpublished = {\url{https://www.ipcc.ch/2022/04/04/ipcc-ar6-wgiii-pressrelease/}},
month = {},
year = {2022},
note = {(Accessed on 09/04/2023)}
}

@article{papoudakis2019dealing,
  title={Dealing with non-stationarity in multi-agent deep reinforcement learning},
  author={Papoudakis, Georgios and Christianos, Filippos and Rahman, Arrasy and Albrecht, Stefano V},
  journal={arXiv preprint arXiv:1906.04737},
  year={2019}
}

@article{kittel2021lakes,
  title={From lakes and glades to viability algorithms: automatic classification of system states according to the topology of sustainable management},
  author={Kittel, Tim and M{\"u}ller-Hansen, Finn and Koch, Rebekka and Heitzig, Jobst and Deffuant, Guillaume and Mathias, Jean-Denis and Kurths, J{\"u}rgen},
  journal={The European Physical Journal Special Topics},
  volume={230},
  pages={3133--3152},
  year={2021},
  publisher={Springer}
}

@article{nordhaus2015climate,
  title={Climate clubs: Overcoming free-riding in international climate policy},
  author={Nordhaus, William},
  journal={American Economic Review},
  volume={105},
  number={4},
  pages={1339--1370},
  year={2015},
  publisher={American Economic Association 2014 Broadway, Suite 305, Nashville, TN 37203}
}

@misc{Integrat69:online,
author = {UN},
title = {Integrated Assessment Models (IAMs) and Energy-Environment-Economy (E3) models | UNFCCC},
howpublished = {\url{https://unfccc.int/topics/mitigation/workstreams/response-measures/modelling-tools-to-assess-the-impact-of-the-implementation-of-response-measures/integrated-assessment-models-iams-and-energy-environment-economy-e3-models}},
month = {},
year = {2023},
note = {(Accessed on 09/22/2023)}
}

@article{farmer2015third,
  title={A third wave in the economics of climate change},
  author={Farmer, J Doyne and Hepburn, Cameron and Mealy, Penny and Teytelboym, Alexander},
  journal={Environmental and Resource Economics},
  volume={62},
  pages={329--357},
  year={2015},
  publisher={Springer}
}

@article{stone2008global,
  title={Global public policy, transnational policy communities, and their networks},
  author={Stone, Diane},
  journal={Policy studies journal},
  volume={36},
  number={1},
  pages={19--38},
  year={2008},
  publisher={Wiley Online Library}
}

@article{yu2022surprising,
  title={The surprising effectiveness of ppo in cooperative multi-agent games},
  author={Yu, Chao and Velu, Akash and Vinitsky, Eugene and Gao, Jiaxuan and Wang, Yu and Bayen, Alexandre and Wu, Yi},
  journal={Advances in Neural Information Processing Systems},
  volume={35},
  pages={24611--24624},
  year={2022}
}

@article{schulman2017proximal,
  title={Proximal policy optimization algorithms},
  author={Schulman, John and Wolski, Filip and Dhariwal, Prafulla and Radford, Alec and Klimov, Oleg},
  journal={arXiv preprint arXiv:1707.06347},
  year={2017}
}

@article{kelly1999integrated,
  title={Integrated assessment models for climate change control},
  author={Kelly, David L and Kolstad, Charles D and others},
  journal={International yearbook of environmental and resource economics},
  volume={2000},
  pages={171--197},
  year={1999},
  publisher={Edward Elgar Publishing Ltd Northampton}
}

@article{van2019improved,
  title={Improved modelling of lifestyle changes in Integrated Assessment Models: Cross-disciplinary insights from methodologies and theories},
  author={van den Berg, Nicole J and Hof, Andries F and Akenji, Lewis and Edelenbosch, Oreane Y and van Sluisveld, Mari{\"e}sse AE and Timmer, Vanessa J and van Vuuren, Detlef P},
  journal={Energy Strategy Reviews},
  volume={26},
  pages={100420},
  year={2019},
  publisher={Elsevier}
}

@article{sert2020segregation,
  title={Segregation dynamics with reinforcement learning and agent based modeling},
  author={Sert, Egemen and Bar-Yam, Yaneer and Morales, Alfredo J},
  journal={Scientific reports},
  volume={10},
  number={1},
  pages={11771},
  year={2020},
  publisher={Nature Publishing Group UK London}
}

@article{liang2020agent,
  title={Agent-based modeling in electricity market using deep deterministic policy gradient algorithm},
  author={Liang, Yanchang and Guo, Chunlin and Ding, Zhaohao and Hua, Huichun},
  journal={IEEE transactions on power systems},
  volume={35},
  number={6},
  pages={4180--4192},
  year={2020},
  publisher={IEEE}
}

@book{cairney2016politics,
  title={The politics of evidence-based policy making},
  author={Cairney, Paul},
  year={2016},
  publisher={Springer}
}

@inproceedings{Huang2018,
  title={{Establishing Appropriate Trust via Critical States}},
  author={Huang, Sandy H and Bhatia, Kush and Abbeel, Pieter and Dragan, Anca D},
  booktitle={2018 IEEE/RSJ International Conference on Intelligent Robots and Systems (IROS)},
  pages={3929--3936},
  year={2018},
  organization={IEEE}
}

@article{skalse2022defining,
  title={Defining and characterizing reward gaming},
  author={Skalse, Joar and Howe, Nikolaus and Krasheninnikov, Dmitrii and Krueger, David},
  journal={Advances in Neural Information Processing Systems},
  volume={35},
  pages={9460--9471},
  year={2022}
}

@article{laidlaw2024preventing,
  title={Preventing reward hacking with occupancy measure regularization},
  author={Laidlaw, Cassidy and Singhal, Shivam and Dragan, Anca},
  journal={arXiv preprint arXiv:2403.03185},
  year={2024}
}

@article{albrecht2018autonomous,
  title={Autonomous agents modelling other agents: A comprehensive survey and open problems},
  author={Albrecht, Stefano V and Stone, Peter},
  journal={Artificial Intelligence},
  volume={258},
  pages={66--95},
  year={2018},
  publisher={Elsevier}
}

@article{patterson2023backlash,
  title={Backlash to climate policy},
  author={Patterson, James J},
  journal={Global Environmental Politics},
  volume={23},
  number={1},
  pages={68--90},
  year={2023},
  publisher={MIT Press One Broadway, 12th Floor, Cambridge, Massachusetts 02142, USA~…}
}

@article{bauer2021digital,
  title={A digital twin of Earth for the green transition},
  author={Bauer, Peter and Stevens, Bjorn and Hazeleger, Wilco},
  journal={Nature Climate Change},
  volume={11},
  number={2},
  pages={80--83},
  year={2021},
  publisher={Nature Publishing Group}
}

@article{shapley1953stochastic,
  title={Stochastic games},
  author={Shapley, Lloyd S},
  journal={Proceedings of the national academy of sciences},
  volume={39},
  number={10},
  pages={1095--1100},
  year={1953},
  publisher={National Acad Sciences}
}

@inproceedings{hansen2004dynamic,
  title={Dynamic programming for partially observable stochastic games},
  author={Hansen, Eric A and Bernstein, Daniel S and Zilberstein, Shlomo},
  booktitle={AAAI},
  volume={4},
  pages={709--715},
  year={2004}
}

@inproceedings{christianos2021scaling,
  title={Scaling multi-agent reinforcement learning with selective parameter sharing},
  author={Christianos, Filippos and Papoudakis, Georgios and Rahman, Muhammad A and Albrecht, Stefano V},
  booktitle={International Conference on Machine Learning},
  pages={1989--1998},
  year={2021},
  organization={PMLR}
}

@article{van2020anticipating,
  title={Anticipating futures through models: the rise of Integrated Assessment Modelling in the climate science-policy interface since 1970},
  author={Van Beek, Lisette and Hajer, Maarten and Pelzer, Peter and van Vuuren, Detlef and Cassen, Christophe},
  journal={Global Environmental Change},
  volume={65},
  pages={102191},
  year={2020},
  publisher={Elsevier}
}

@article{bernard2008gemini,
  title={GEMINI-E3, a general equilibrium model of international--national interactions between economy, energy and the environment},
  author={Bernard, Alain and Vielle, Marc},
  journal={Computational Management Science},
  volume={5},
  number={3},
  pages={173--206},
  year={2008},
  publisher={Springer}
}

@article{van2023multimodel,
  title={A multimodel analysis of post-Glasgow climate targets and feasibility challenges},
  author={van de Ven, Dirk-Jan and Mittal, Shivika and Gambhir, Ajay and Lamboll, Robin D and Doukas, Haris and Giarola, Sara and Hawkes, Adam and Koasidis, Konstantinos and K{\"o}berle, Alexandre C and McJeon, Haewon and others},
  journal={Nature Climate Change},
  volume={13},
  number={6},
  pages={570--578},
  year={2023},
  publisher={Nature Publishing Group UK London}
}

@article{gambhir2019review,
  title={A review of criticisms of integrated assessment models and proposed approaches to address these, through the lens of BECCS},
  author={Gambhir, Ajay and Butnar, Isabela and Li, Pei-Hao and Smith, Pete and Strachan, Neil},
  journal={Energies},
  volume={12},
  number={9},
  pages={1747},
  year={2019},
  publisher={MDPI}
}

@inproceedings{Islam2020,
 title={{Towards Quantification of Explainability in Explainable Artificial Intelligence Methods}},
  author={Islam, Sheikh Rabiul and Eberle, William and Ghafoor, Sheikh K},
  booktitle={The thirty-third international flairs conference},
  year={2020}
}

@article{Rudin2019,
  title={{Stop explaining black box machine learning models for high stakes decisions and use interpretable models instead}},
  author={Rudin, Cynthia},
  journal={Nature machine intelligence},
  volume={1},
  number={5},
  pages={206--215},
  year={2019},
  publisher={Nature Publishing Group UK London}
}

@inproceedings{Li2018,
  title={{Deep learning for case-based reasoning through prototypes: A neural network that explains its predictions}},
  author={Li, Oscar and Liu, Hao and Chen, Chaofan and Rudin, Cynthia},
  booktitle={Proceedings of the AAAI Conference on Artificial Intelligence},
  volume={32},
  number={1},
  year={2018}
}

@article{Adadi2018,
 title={{Peeking Inside the Black-Box: A Survey on Explainable Artificial Intelligence (XAI)}},
  author={Adadi, Amina and Berrada, Mohammed},
  journal={IEEE access},
  volume={6},
  pages={52138--52160},
  year={2018},
  publisher={IEEE}
}

@article{Lipton2018,
 title={{The Mythos of Model Interpretability}},
  author={Lipton, Zachary C},
  journal={Queue},
  volume={16},
  number={3},
  pages={31--57},
  year={2018},
  publisher={ACM New York, NY, USA}
}

@article{Glanois2021,
 title={{A Survey on Interpretable Reinforcement Learning}},
  author={Glanois, Claire and Weng, Paul and Zimmer, Matthieu and Li, Dong and Yang, Tianpei and Hao, Jianye and Liu, Wulong},
  journal={arXiv preprint arXiv:2112.13112},
  year={2021}
}

@article{axtell2022agent,
  title={Agent-based modeling in economics and finance: Past, present, and future},
  author={Axtell, Robert L and Farmer, J Doyne},
  journal={Journal of Economic Literature},
  pages={1--101},
  year={2022},
  publisher={American Economic Association}
}

@inproceedings{wise2017transportation,
  title={Transportation in agent-based urban modelling},
  author={Wise, Sarah and Crooks, Andrew and Batty, Michael},
  booktitle={Agent Based Modelling of Urban Systems: First International Workshop, ABMUS 2016, Held in Conjunction with AAMAS, Singapore, Singapore, May 10, 2016, Revised, Selected, and Invited Papers 1},
  pages={129--148},
  year={2017},
  organization={Springer}
}

@article{giarola2022muse,
  title={MUSE: An open-source agent-based integrated assessment modelling framework},
  author={Giarola, Sara and Sachs, Julia and d’Avezac, Mayeul and Kell, Alexander and Hawkes, Adam},
  journal={Energy Strategy Reviews},
  volume={44},
  pages={100964},
  year={2022},
  publisher={Elsevier}
}

@article{heuillet2021explainability,
  title={Explainability in deep reinforcement learning},
  author={Heuillet, Alexandre and Couthouis, Fabien and D{\'\i}az-Rodr{\'\i}guez, Natalia},
  journal={Knowledge-Based Systems},
  volume={214},
  pages={106685},
  year={2021},
  publisher={Elsevier}
}

@misc{bradbury2018jax,
  title={JAX: composable transformations of Python+ NumPy programs},
  author={Bradbury, James and Frostig, Roy and Hawkins, Peter and Johnson, Matthew James and Leary, Chris and Maclaurin, Dougal and Necula, George and Paszke, Adam and VanderPlas, Jake and Wanderman-Milne, Skye and others},
  year={2018}
}

@article{dearing2014safe,
  title={Safe and just operating spaces for regional social-ecological systems},
  author={Dearing, John A and Wang, Rong and Zhang, Ke and Dyke, James G and Haberl, Helmut and Hossain, Md Sarwar and Langdon, Peter G and Lenton, Timothy M and Raworth, Kate and Brown, Sally and others},
  journal={Global Environmental Change},
  volume={28},
  pages={227--238},
  year={2014},
  publisher={Elsevier}
}

@article{hussein2017imitation,
  title={Imitation learning: A survey of learning methods},
  author={Hussein, Ahmed and Gaber, Mohamed Medhat and Elyan, Eyad and Jayne, Chrisina},
  journal={ACM Computing Surveys (CSUR)},
  volume={50},
  number={2},
  pages={1--35},
  year={2017},
  publisher={ACM New York, NY, USA}
}

@article{kellett2019feedback,
  title={Feedback, dynamics, and optimal control in climate economics},
  author={Kellett, Christopher M and Weller, Steven R and Faulwasser, Timm and Gr{\"u}ne, Lars and Semmler, Willi},
  journal={Annual Reviews in Control},
  volume={47},
  pages={7--20},
  year={2019},
  publisher={Elsevier}
}

@article{nordhaus2010economic,
  title={Economic aspects of global warming in a post-Copenhagen environment},
  author={Nordhaus, William D},
  journal={Proceedings of the National Academy of Sciences},
  volume={107},
  number={26},
  pages={11721--11726},
  year={2010},
  publisher={National Acad Sciences}
}

@article{garcia1989MPC,
  title={Model predictive control: Theory and practice—A survey},
  author={Garcia, Carlos E and Prett, David M and Morari, Manfred},
  journal={Automatica},
  volume={25},
  number={3},
  pages={335--348},
  year={1989},
  publisher={Elsevier}
}

@article{dellink2019sectoral,
  title={The sectoral and regional economic consequences of climate change to 2060},
  author={Dellink, Rob and Lanzi, Elisa and Chateau, Jean},
  journal={Environmental and resource economics},
  volume={72},
  pages={309--363},
  year={2019},
  publisher={Springer}
}

@inproceedings{lu2022model,
  title={Model-free opponent shaping},
  author={Lu, Christopher and Willi, Timon and De Witt, Christian A Schroeder and Foerster, Jakob},
  booktitle={International Conference on Machine Learning},
  pages={14398--14411},
  year={2022},
  organization={PMLR}
}

@article{ma2024efficient,
  title={Efficient and scalable reinforcement learning for large-scale network control},
  author={Ma, Chengdong and Li, Aming and Du, Yali and Dong, Hao and Yang, Yaodong},
  journal={Nature Machine Intelligence},
  pages={1--15},
  year={2024},
  publisher={Nature Publishing Group UK London}
}

@inproceedings{nayak2023scalable,
  title={Scalable multi-agent reinforcement learning through intelligent information aggregation},
  author={Nayak, Siddharth and Choi, Kenneth and Ding, Wenqi and Dolan, Sydney and Gopalakrishnan, Karthik and Balakrishnan, Hamsa},
  booktitle={International Conference on Machine Learning},
  pages={25817--25833},
  year={2023},
  organization={PMLR}
}

@article{kazemkhani2024gpudrive,
  title={GPUDrive: Data-driven, multi-agent driving simulation at 1 million FPS},
  author={Kazemkhani, Saman and Pandya, Aarav and Cornelisse, Daphne and Shacklett, Brennan and Vinitsky, Eugene},
  journal={arXiv preprint arXiv:2408.01584},
  year={2024}
}

@article{suarez2019neural,
  title={Neural MMO: A massively multiagent game environment for training and evaluating intelligent agents},
  author={Suarez, Joseph and Du, Yilun and Isola, Phillip and Mordatch, Igor},
  journal={arXiv preprint arXiv:1903.00784},
  year={2019}
}

\end{Backmatter}
 \newpage
\appendix
\section{Further AYS Environment Details}
\label{appendix:further_ays_details}

\begin {table}[h]
\caption{AYS numerical parameters \citep{kittel2021lakes}.}
\label{table:ays_params}
\begin{center}
 \begin{tabular}{|c|l|p{9cm}|} 
    \hline
    \textbf{Parameter}&\textbf{Value}&\textbf{Description}\\[0.8ex] 
    \hline
    $\tau_A$ & 50 Years & Atmospheric carbon decay \\
    $\beta$ & 3\% per Year & Economic growth \\
    $\xi_i$ & $\in (0, 1)$ & Agent specific climate damage \\
    $\theta$ & $8.57 \times 10^{-5}$ & Temperature sensitivity \\
    $\tau_S$ & 50 Years & Renewable knowledge stock decay \\
    $\phi$ & $4.7 \times 10^{10} \; GJGtC^{-1}$ & Fossil fuel combustion efficiency \\
    $\sigma$ & $4 \times 10^{12} \; GJ$ & Break-even renewable knowledge - value at which fossil fuels and renewables have the same cost \\
    $\rho$ & 2 & Renewable knowledge learning rate \\
    $\epsilon$ & $147 \; \$ GJ^{-1}$ & Energy efficiency \\
    \hline
\end{tabular}
\end{center}
\end{table}

\newpage
\section{Hyperparameters}
\label{appendix:rl_training_details}

\begin{table}[H]
    \centering
    \begin{tabular}{|l|c|}
        \hline
         Parameter & Value  \\
         \hline
         RL Algorithm & IPPO \\
         Actor Layers & [128, RNN*, 256, output dim(4)]\\
         Critic Layers & [128, RNN*, 128, output dim(1)] \\
         GRU Hidden Dim & 256 \\
         Clip EPS & 0.2 \\
         Entropy Coefficient & 0.01 \\
         Lambda (for GAE) & 0.95 \\
         Gamma & 0.99 \\
         Learning Rate & $2.5e^{-4,LR}$ \\
         Max Grad Norm & 0.5 \\
         Non Linearity & relu \\
         Number of Minibatches & 4 \\
         Optimiser & adam\\
         Rollout Length & 256 \\
         Seeds & 28, 10, 98, 44, 22, 68 \\
         Update Epochs & 4 \\
         VF Coef & 0.5 \\
         \hline
         
    \end{tabular}
    \caption{Table of training hyperparameters.}
    \label{tab:hyperparam_table}
\end{table}

* shares the same head up until the RNN (GRU aggregator) output then split to actor and critic for further layers. \\
$^{LR}$ with annealed learning rate

\end{document}